\documentclass[11pt]{article}
\usepackage{dcolumn}% Align table columns on decimal point
\usepackage{bm}% bold math

\usepackage[square,numbers,compress,sort]{natbib}
\usepackage{graphicx}% Include figure files
\usepackage{amssymb,amsmath}
\usepackage{multirow}
\usepackage{slashed,xfrac}
\usepackage[makeroom]{cancel}
\usepackage{color,url}
\usepackage[colorlinks=true
,urlcolor=blue
,anchorcolor=blue
,citecolor=blue
,filecolor=blue
,linkcolor=blue
,menucolor=blue
,linktocpage=true
,pdfproducer=medialab
,pdfa=true
]{hyperref}

\usepackage{epsfig,psfrag,rotating,soul}
\usepackage{rotfloat}
\usepackage[normalem]{ulem}

\graphicspath{{figs/}}

\evensidemargin \oddsidemargin
\allowdisplaybreaks

\newcommand{\be}{\begin{equation}}
\newcommand{\ee}{\end{equation}}
\newcommand{\ba}{\begin{eqnarray}}
\newcommand{\ea}{\end{eqnarray}}

\def\thefootnote{\fnsymbol{footnote}}

\begin{document}
\thispagestyle{empty}

\begin{flushright}
IFT-UAM/CSIC-16-137\\
FTUAM-16-46\\
\end{flushright}

\vspace{0.5cm}

\begin{center}

\begin{Large}
\textbf{\textsc{Effective lepton flavor violating $\boldsymbol{H\ell_i\ell_j}$ vertex from right-handed neutrinos within the mass insertion approximation}} 
\end{Large}

\vspace{1cm}

{\sc
E. Arganda$^1$%
\footnote{\tt \href{mailto:ernesto.arganda@fisica.unlp.edu.ar}{ernesto.arganda@fisica.unlp.edu.ar}}%
, M.J. Herrero$^2$%
\footnote{\tt \href{mailto:maria.herrero@uam.es}{maria.herrero@uam.es}}%
, X. Marcano$^2$%
\footnote{\tt \href{mailto:xabier.marcano@uam.es}{xabier.marcano@uam.es}}%
, R. Morales$^1$%
\footnote{\tt \href{mailto:roberto.morales@fisica.unlp.edu.ar}{roberto.morales@fisica.unlp.edu.ar}}%
, A. Szynkman$^1$%
\footnote{\tt \href{mailto:szynkman@fisica.unlp.edu.ar}{szynkman@fisica.unlp.edu.ar}}%
}

\vspace*{.7cm}

{\sl
$^1$IFLP, CONICET - Departamento de F\'isica, Universidad Nacional de La Plata,\\
C.C. 67, 1900 La Plata, Argentina

\vspace*{0.1cm}

$^2$Departamento de F\'{\i}sica Te\'orica and Instituto de F\'{\i}sica Te\'orica, IFT-UAM/CSIC,\\
Universidad Aut\'onoma de Madrid, Cantoblanco, 28049 Madrid, Spain

}

\end{center}

\vspace*{0.1cm}

\begin{abstract}
\noindent
In this work we present a new computation of the lepton flavor violating Higgs boson decays that are generated radiatively to one-loop from heavy right-handed neutrinos. We work within the context of the inverse seesaw model with three $\nu_R$ and three extra singlets $X$, but the results could be generalized to other low scale seesaw models. The novelty of our computation is that it uses a completely different method by means of the mass insertion approximation which works with the electroweak interaction states instead of the usual 9 physical neutrino mass eigenstates of the inverse seesaw model. This method also allows us to write the analytical results explicitly in terms of the most  relevant model parameters, that are the neutrino Yukawa coupling matrix $Y_\nu$ and the right-handed mass matrix $M_R$, which is very convenient for a phenomenological analysis.  This $Y_\nu$ matrix, being generically nondiagonal in flavor space, is the only responsible for the induced charged lepton flavor violating processes of our interest.  
We perform the calculation of the decay amplitude up to order ${\cal O}(Y_\nu^2+Y_\nu^4)$.
We also study numerically the goodness of the mass insertion approximation results.
In the last part we present the computation of the relevant one-loop effective vertex  $H\ell_i\ell_j$ for the lepton flavor violating Higgs decay which is derived from a large $M_R$ mass expansion of the form factors. We believe that our simple formula found for this effective vertex can be of interest  
for other researchers who wish to estimate the $H \to \ell_i \bar \ell_j$  rates in a fast way in terms of their own preferred input values for  
the relevant model parameters $Y_\nu$ and $M_R$. 
 \end{abstract}

\def\thefootnote{\arabic{footnote}}
\setcounter{page}{0}
\setcounter{footnote}{0}

\newpage
%%%%%%%%%%%%%%%%%%%%%%%%%%%%%%%%%%%%%
\section{Introduction}
\label{intro}
 The study of lepton flavor violating (LFV) processes provides undoubtedly one of the most promising avenues to explore the existence of new physics beyond the standard model (SM) of particle physics. Lepton flavor symmetry is indeed an exact symmetry of the SM and therefore it predicts vanishing rates for all these LFV processes to all orders in perturbation theory.  Interestingly, any experimental signal of LFV will in consequence indicate that some new physics, either new particles or new interactions must be the responsible for it. The SM, on the other hand, has to be modified in any case in order to include the observed neutrino masses and oscillations, and for this purpose it seems quite natural to extend it with the addition of right-handed neutrinos which, in contrast to the other right-handed leptons and quarks, were ignored  in its construction.  
 
 The charged LFV processes are particularly adequate to study one of the most important indirect effects that are derived from the existence of the right-handed neutrinos. This occurs because the right-handed neutrinos carry lepton flavor number and can interact with the left-handed neutrinos of different flavor via their Yukawa couplings which may be described by flavor nondiagonal matrices. Thus,  the right-handed neutrinos may contribute to LFV processes via their radiative corrections to the observables that describe these processes. Specially, in the context of low scale seesaw models where the mass scale associated to the right-handed neutrinos $M_R$ could be not so far above from the electroweak (EW) scale and even be accessible to the present colliders like the CERN Large Hadron Collider (LHC)  if they are close to the TeV scale. 
Charged LFV processes within low scale seesaw models have been extensively studied in the literature \cite{Ilakovac:1994kj, Abada:2012cq, Alonso:2012ji, Abada:2012mc, Abada:2013aba, Abada:2014kba, Abada:2014cca, Abada:2015zea, Arganda:2014dta, Arganda:2015naa, Arganda:2015ija, Abada:2015oba, DeRomeri:2016gum, Abada:2016plb, Abada:2016vzu}.
 
Here we are interested in the study of one of these processes, the LFV Higgs boson decays (LFVHD) into charged leptons with different flavor, $H \to \ell_i \bar \ell_j$ with $i \neq j$.  These particular decays are at present being searched for very actively in the LHC experiments and, in fact, there are already significant bounds set from the absence of signals in both ATLAS and CMS~\cite{Khachatryan:2015kon,Aad:2015gha,CMS:2015udp,Aad:2016blu,CMS:2016qvi}. The present $95\%$ CL bounds are
\begin{align}
{\rm BR}(H\to\mu e) &< 3.6\times10^{-4}\,,\\
 {\rm BR}(H\to\tau e) &< 0.70\times10^{-2}\,,\\
 {\rm BR}(H\to\tau\mu ) &< 1.20\times10^{-2}\,.
 \end{align}
We are motivated in particular to the study of these LFVHD under the hypothesis that they are originated to one-loop level from the radiative corrections of right-handed neutrinos within low scale seesaw models.  Concretely, we choose to work in the context of the inverse seesaw (ISS) model~\cite{Bernabeu:1987gr,Mohapatra:1986aw,Mohapatra:1986bd,GonzalezGarcia:1991be} with three right-handed neutrinos and three extra singlets of opposite lepton number. These  LFVHD  have been extensively studied in the literature in the context of seesaw models (both High scale and Low scale seesaw models) and there are significant predictions for their rates as a function of the various model parameters.  In particular, the partial widths $\Gamma(H \to \ell_i \bar \ell_j)$  were first computed by a diagrammatic procedure to one-loop in a generic seesaw model  with three generations of right-handed neutrinos  in \cite{Pilaftsis:1992st} and in \cite{Arganda:2004bz}, and also in the ISS model that we are interested in with three right-handed neutrinos and three extra singlets in \cite{Arganda:2014dta}. All these computations were performed in the physical neutrino mass basis, and the results have been provided mainly in terms of the physical mass parameters and the rotation matrix entries, which connect the initial electroweak interaction basis with the final mass basis.

In this work, we will perform a completely different and independent analysis of these LFVHD rates within the ISS model. Instead of using the physical neutrino basis, which amounts to a  heavy numerical computation of the full set of diagrams with all the 9 physical neutrinos in the loops, and considering the complex dependence on the ISS model parameters hidden in the values of the rotated physical states couplings, we will perform our computation of the LFVHD widths directly in the chiral electroweak interaction basis with left- and right-handed neutrinos being the fields propagating in the loops, and we will express the result explicitly in terms of the model parameters, most relevantly,  the $3 \times 3$ neutrino Yukawa coupling matrix $Y_\nu$ and the right-handed mass matrix $M_R$.  We will present, for the first time to our knowledge, this one-loop computation done in the mass insertion approximation (MIA) which turns out to be a very powerful tool to use in this context of low scale seesaw models with heavy but not extremely heavy right-handed neutrinos.  Another different use of the MIA was previously done in ~\cite{Arganda:2015uca} for the computation of the LFVHD rates but in the different context of supersymmetric neutrino models where the LFV was induced differently from the soft SUSY breaking mass insertions changing lepton flavor. Here, in contrast, the MIA provides the results in terms of a well defined expansion in powers of $Y_\nu$, which is the unique relevant origin of  lepton flavor violation in the present work, and therefore it is a very useful and convenient method for an easier and clearer interpretation of the related phenomenology.  For the present study of the $H \to \ell_i \bar \ell_j$ decay amplitude  we will calculate this MIA expansion first to leading order, ${\cal O}((Y_\nu Y_\nu^\dagger)_{ij})$, and second to the next to leading order, i.e. including terms up to ${\cal O}((Y_\nu Y_\nu^\dagger Y_\nu Y_\nu^\dagger)_{ij})$, and we will explore the goodness of this approximation. In addition to the computation of the form factors involved in these LFVHD  we will also calculate with the MIA the one-loop effective vertex ${H\ell_i\ell_j}$ that is the relevant one for these decays. In getting this effective vertex we will explore the proper large $M_R$ mass expansion, which in the present case must apply for the assumed mass hierarchy, $m_{\ell_{i,j}}\ll vY_\nu,m_W,m_H\ll M_R$ with $m_{\ell_{i,j}}$ the lepton masses, $v$ the Higgs vacuum expectation value, and $m_W$, $m_H$, the $W$ boson and Higgs particle masses.   The most appealing feature of our computation is that it provides very simple formulas, which seem to work very well, for both the one-loop effective ${H\ell_i\ell_j}$ LFV vertex  and the partial width  $\Gamma(H \to \ell_i \bar\ell_j)$  in terms of the most relevant parameters, mainly $Y_\nu$ and $M_R$. These simple formulas could be easily used by other authors to estimate rapidly the LFVHD rates with their own inputs for $Y_\nu$ and $M_R$ and without the need of a heavy numerical computation.  

The paper is organized as follows. The first section is devoted to summarizing the main features of the ISS model in terms of the EW interaction basis and derives the set of relevant Feynman rules that are needed for the present MIA computation. The second section presents the computation of ${\Gamma(H\to\ell_i\bar\ell_j)}$ to one-loop within the MIA and explores the goodness of this approximation by comparing the MIA and the full one-loop results. In the third section we present the computation of the one-loop effective vertex for LFVHD. The summary of our main findings is given in the conclusions section. The technical details of the present computation, the complete set of formulas for the form factors in both the Feynman-'t Hooft gauge and the unitary gauge, and the large $M_R$ expansion results are collected in the appendices.

%%%%%%%%%%%%%%%%%%%%%%%%%%%%%%%%%%%%%
\section{The proper $\boldsymbol{\nu_R}$ basis and Feynman rules for a MIA computation}
\label{model} 

For the final purpose in this work of computing the one-loop generated effective  ${H\ell_i\ell_j}$ vertex from right-handed neutrinos within the MIA, it is important first to choose the proper EW interaction basis, i.e.,  the basis for the left- and right-handed chiral neutrino fields, and to set up the necessary Feynman rules in terms of these fields. The basic points of the MIA and its simplicity are precisely based on the use of the EW basis instead of the mass basis which is the one usually used in the literature for the one-loop generated LFV observables in models with massive Majorana neutrinos. 
As we have said, we work in the context of low scale seesaw models, and perform the analytical computation of the LFV Higgs form factors and effective vertices for one of their particular realizations, the ISS model, although as we will see later, the results could be  generalized to other low scale seesaw models.
Our chiral fields for the present computation will be therefore the left- and right-handed neutrinos  of the ISS. In this section we present our choice for the proper chiral basis and Feynman rules of the ISS, set the relevant input model parameters and prepare the set up of the model for the computation of the one-loop LFV Higgs form factors and effective vertices in the next sections. 
  
Regarding the general features of low scale seesaw models that are relevant for the present work, it is important to remind that they assume approximate symmetries, as it is the case of an approximate lepton number (LN) conservation, in order to explain the lightness of the observed neutrino masses.
A particular example of such a model is the  ISS model, which assumes that there is a $U(1)_L$ symmetry which is broken only by a small parameter $\mu_X$. Since setting this parameter to zero would increase the symmetry of the model, it is natural to consider it to be small. 
Then, one can explain the smallness of the neutrino masses by relating them to this small parameter $\mu_X$. 
Contrary to the Type-I seesaw model, where the lightness of $m_\nu$ is explained by the small ratio between two very distant scales, that of the EW symmetry breaking, given by the Higgs vacuum expectation value $v = 174$~GeV, divided by a large scale, given by the Majorana mass associated to the LN breaking which could be as heavy as $10^{14-15}$ GeV; in the ISS instead $m_\nu$ is considered to be proportional to the small LN breaking scale, $\mu_X$. This is why it is known as the {\it inverse} seesaw.
Interestingly, the introduction of this small scale $\mu_X$ allows us both to accommodate successfully the light neutrino data and to incorporate the new moderately heavy neutrinos, say at the TeV scale, with potentially large Yukawa couplings, say with $Y_\nu^2/4\pi \lesssim \mathcal O(1)$, which could have relevant implications for phenomenology. 

Concretely, the ISS that we consider in our work includes pairs of fermionic singlets, $(\nu_R,X)$, with opposite LN and assumes that the LN is only violated by the {\it naturally} small $\mu_X$ Majorana mass term for the singlets $X$. 
In order to accommodate light neutrino data, one needs to add more than one pair of fermionic singlets. Following the SM pattern with three fermion generations, we consider here adding three of these extra pairs to the SM particle content. 
Therefore, our starting Lagrangian of the ISS reads as follows:
\begin{equation}
 \label{ISSlagrangian}
 \mathcal{L}_\mathrm{ISS} = - Y^{ij}_\nu \overline{L_{i}} \widetilde{\Phi} \nu_{Rj} - M_R^{ij} \overline{\nu_{Ri}^c} X_j - \frac{1}{2} \mu_{X}^{ij} \overline{X_{i}^c} X_{j} + H.c.\,,
\end{equation}
where $L$ is the SM lepton doublet, $\widetilde{\Phi}=i\sigma_2\Phi^*$ with $\Phi$ the SM Higgs doublet and 
$i,j$ are indices in flavor space that run from 1 to 3. Correspondingly, $Y_\nu$, $\mu_{X}$ and $M_R$ are $3\times 3$ matrices.
The $C$-conjugate fermion fields are defined here as $f_L^c=(f_L)^c=(f^c)_R$ and $f_R^c=(f_R)^c=(f^c)_L$.

After the electroweak symmetry breaking, one gets the complete $9\times9$ neutrino mass matrix of the ISS that, in the electroweak interaction basis $(\nu_L^c\,,\;\nu_R\,,\;X)$, reads: 
\begin{equation}
\label{ISSmatrix}
 M_{\mathrm{ISS}}=\left(\begin{array}{c c c} 0 & m_D & 0 \\ m_D^T & 0 & M_R \\ 0 & M_R^T & \mu_X \end{array}\right)\,,
\end{equation}
with $m_D=v Y_\nu$, and $v = 174\,\mathrm{GeV}$. This mass matrix can be diagonalized using a $9\times9$ unitary matrix $U_\nu$ according to:
\begin{equation}
U_\nu^T M_{\rm ISS} U_\nu = {\rm diag}(m_{n_1},\dots,m_{n_9}),
\end{equation}
and leading to 9 physical neutrino mass eigenstates $n_i$, $i=1,.,9$, which are Majorana fermions, i.e. they are their own antiparticles. 
Then, the relation between the EW chiral and physical neutrino eigenstates is given by 
\begin{equation}\label{EWtoPhysical}
\left(\begin{array}{c} \nu_L^c \\ \nu_R \\ X \end{array}\right) = U_\nu P_R \left(\begin{array}{c} n_1\\ \vdots \\ n_9\end{array}\right),
\quad
\left(\begin{array}{c} \nu_L \\ \nu_R^c \\ X^c \end{array}\right) = U_\nu^* P_L \left(\begin{array}{c} n_1\\ \vdots \\ n_9\end{array}\right).
\end{equation}

In the case $\mu_X \ll m_D \ll M_R$, it is possible to diagonalize $M_{\rm ISS}$ by blocks, leading to the following $3 \times 3$ light neutrino mass matrix
\begin{equation}
M_{\mathrm{light}} \simeq m_D {M_R^T}^{-1} \mu_X M_R^{-1} m_D^T\,,
\end{equation}
which is diagonalized by the matrix $U_{\rm PMNS}$:
\begin{equation} \label{mnulight}
 U_{\rm PMNS}^T M_{\mathrm{light}} U_{\rm PMNS} =m_\nu^{\rm diag}\,,
\end{equation}
where $m_\nu^{\rm diag}=\mathrm{diag}(m_{\nu_1}\,, m_{\nu_2}\,, m_{\nu_3})$ is the diagonal matrix that contains the masses of the three lightest neutrinos. Regarding the heavy neutrinos $6 \times 6$ sub-block, its diagonalization in this $\mu_X \ll m_D \ll M_R$ case leads to physical heavy neutrino masses close to the ``big mass'', $m_N  \simeq M_R$.
Besides, it is well known that neutrino oscillation data can be easily accommodated in the model by means of the $\mu_X$ parametrization~\cite{Arganda:2014dta}:
\begin{equation}\label{MUXparam}
\mu_X=M_R^T m_D^{-1} U^*_{\rm PMNS} m_\nu^{\rm diag} U_{\rm PMNS}^\dagger m_D^{T-1} M_R,
\end{equation}
which is easily derived from the previous equations. The advantage of using this parametrization is that we can consider the Yukawa couplings $Y_\nu$ and the $M_R$ masses, which are the most relevant neutrino parameters for LFV processes, as independent input parameters and at the same time be sure that our model fits the light neutrino data. On the other hand, it is also important to remind that there are  three different scales which play different roles in this ISS model: $\mu_X$ controls the smallness of the light neutrino masses, $M_R$ the masses of the new heavy neutrinos, and $m_D=vY_\nu$ involves the interactions between the left- and right-handed neutrinos with the Higgs boson, which are in general nondiagonal in flavor space.
Since they are independent, we can have at the same time large Yukawa couplings, $Y_\nu^2/4\pi\sim\mathcal O(1)$, and  moderate heavy neutrino masses, say with $M_R$ at the TeV range, i.e, reachable at present experiments. These two properties make the ISS an interesting model with a rich phenomenology and important consequences for LFV.  
 
Within the context of the MIA that we are interested in, the previous $9\times9$ mass matrix provides all the relevant masses for the EW eigenstates  and mass insertions that are needed for the present computation. These mass insertions connect two different neutrino states, they are in general flavor nondiagonal, and can be expressed in terms of the three $3 \times 3$ matrices involved, $m_D$, $\mu_X$ and $M_R$. Specifically, the mass insertion given by $m_D$ connects $\nu_L$ and $\nu_R$, $M_R$ connects $\nu_R$ and $X$, and $\mu_X$ connects two  $X$. To simplify the computation, we will use 
the freedom of redefining the new fields ($\nu_R,X$) in such a way that the matrix $M_R$ is flavor diagonal. Thus, all the flavor violation is contained in the matrices $\mu_X$ and $m_D$. Since we are working with $\mu_X$ being extremely  small as to accommodate the light neutrino masses, this mass matrix will be irrelevant for the LFV physics that we will study in this work. Therefore, the only relevant flavor violating insertion will be provided by the $m_D$ matrix and, in consequence, by  the Yukawa coupling matrix $Y_\nu$. 
Regarding the diagonal matrix $M_R$ we will further simplify our computation by considering degenerate diagonal entries, i.e, $M_{R_i}\equiv M_R$. The generalization to the nondegenerate case will be commented in the appendices. 

On the other hand, it should be noticed that the flavor preserving
mass insertions given by $M_R$ can be very large if $M_R$ is taken to be heavy, as it will be our case with $M_R$ being at the TeV scale.   
Since we are finally interested in a perturbative MIA computation of the one-loop LFV Higgs form factors and effective vertices that are valid for heavy $M_R$ masses, we find convenient to use a different chiral basis where the ``big insertions'' given by  $M_R$ are resumed in such a way that the ``large mass'' $M_R$  appears effectively in the denominator of the propagators of the new states.
The key point in choosing this proper chiral basis is provided by the fact that for the quantities of our interest in this work, having $H$, $\ell_i$ and $\ell_j$ in the external particles, the only neutrino states that interact with them are $\nu_L$ and $\nu_R$. The $X$ singlet fields interact exclusively with the $\nu_R$ fields via the $M_R$ mass insertions and, therefore, they will only appear in the computation of the loop diagrams for LFV as internal intermediate states inside internal lines that start and end with $\nu_R$'s. This motivates clearly our choice of modified propagators for the $\nu_R$ fields which are built on purpose to include inside all the effects of the sequential insertions of the $X$ fields,  given each of these insertions by $M_R$. 
More concretely, we sum all the $M_R$ insertions and define two types of  modified propagators: one with the same initial and final particle, corresponding to an even number of $M_R$ mass insertions which we call {\it fat propagators}, and one with different initial and final particles, corresponding to an odd number of insertions. The {\it fat propagator},  that propagates a $\nu_R$ into a $\nu_R$ and contains the sum of all the infinite series of even number of $M_R$ insertions due to the interactions with $X$, is the one we need for the present computation. 
The details of the procedure to reach this proper chiral basis and the derivation of the modified propagators are explained in Appendix~\ref{Propagators}.  
Similar results are obtained within the context of the flavor expansion theorem\footnote{We warmly thank Michael Paraskevas for his kind comment about the similarities between our {\it fat propagators} and the results in the flavor expansion theorem.} \cite{Dedes:2015twa,Rosiek:2015jua}.

\begin{figure}[t!]
\begin{center}
\includegraphics[scale=1]{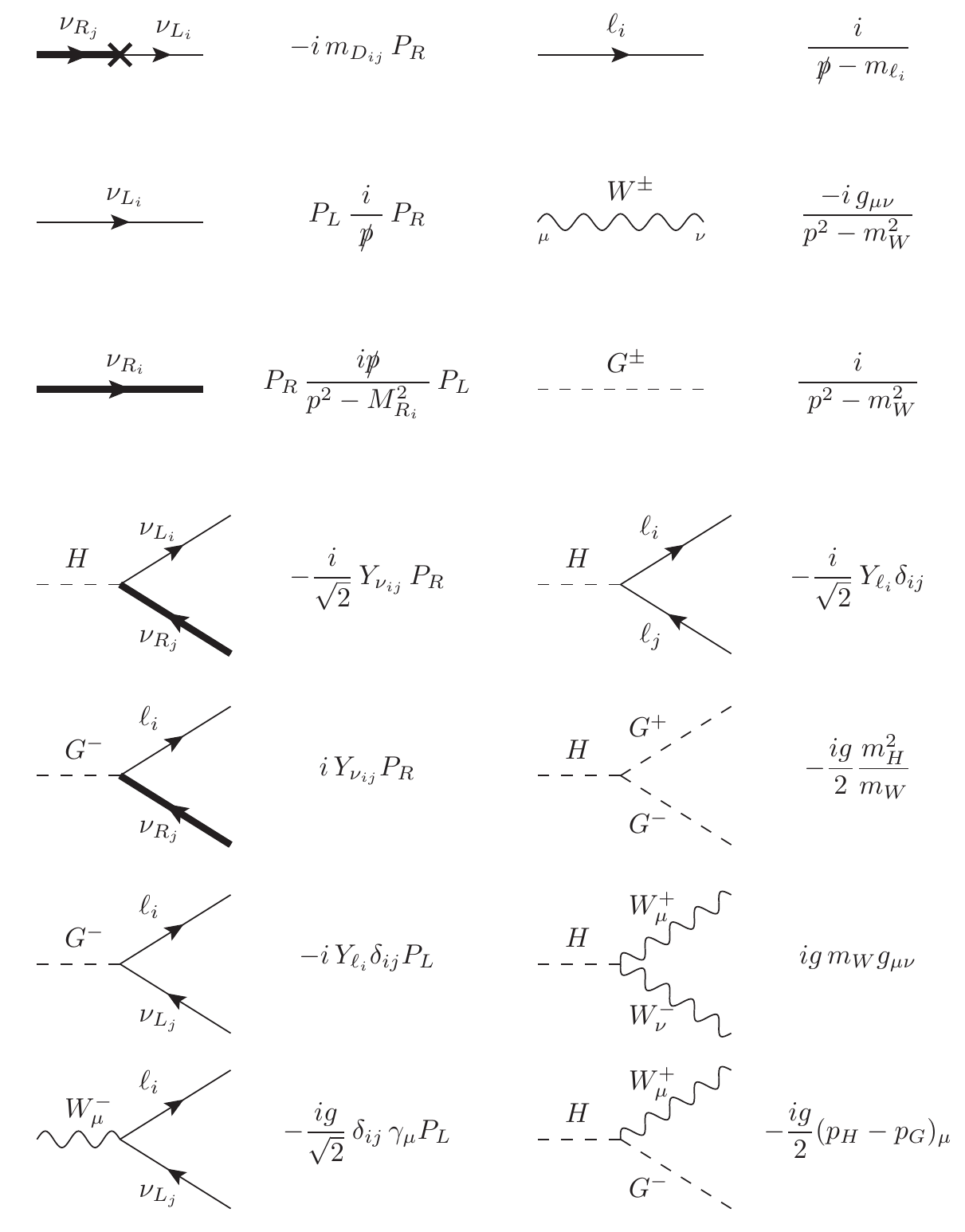}
\caption{Relevant Feynman rules for the present MIA computation of $\Gamma(H\to\ell_k\bar\ell_m)$. The rules involving neutrinos are written in terms of the proper EW chiral basis for $\nu_R$ and $\nu_L$, as defined in Section 2. Some additional SM Feynman rules that are needed are also included, for completeness. Here the Feynman-t'Hooft gauge is selected. The momenta $p_H$ and $p_G$ are incoming.}\label{FR-MIA}
\end{center}
\end{figure}

In order to complete the setup for our computations, we summarize the relevant Feynman rules in our previously chosen proper chiral basis in figure~\ref{FR-MIA}. These include the relevant flavor changing mass insertions, given by $m_D$, the relevant propagators, both the usual SM EW propagators and the new {\it fat propagators} of the $\nu_R$'s, and the relevant interaction vertices, both the SM EW vertices and the new ones involving the $\nu_R$'s. 

Finally, we want to point out that, although we have considered the ISS model to make our computations, our results could be applied in practice to any low scale seesaw model that leads to the same Feynman rules as in figure \ref{FR-MIA}. These are indeed quite generic Feynman rules in models with right-handed heavy neutrinos. The few specific requirements are that the only relevant LFV source is the Yukawa neutrino coupling matrix and that the heavy right-handed neutrino propagator is like our {\it fat propagator} introduced above. 
 
%%%%%%%%%%%%%%%%%%%%%%%%%%%%%%%%%%%%%
\section{$\boldsymbol{\Gamma(H\to\ell_k\bar\ell_m)}$ to one-loop within the MIA}
\label{computationwidth}

Here we present our computation of the partial decay widths for the LFVHD within the ISS model, 
$\Gamma(H \to \ell_k \bar \ell_m)$, with $k,m= 1,2,3$ and $k \neq m$, and explain the details of how we implement the MIA, both to leading order and to next to leading order in the proper expansion in powers of $m_D$ (or equivalently in powers of $Y_\nu$) which, as we have said in the previous section, is the relevant mass insertion producing the needed change of lepton flavor in our observable. In the following, we ignore the potential effects from $\mu_X$, which are expected to be extremely tiny in this flavor changing observable. We first present the decay amplitude and the partial width in terms of the relevant form factors and then we explain the MIA expansion for these form factors. 

The decay amplitude of the process $H(p_1) \to \ell_k(-p_2) \bar \ell_m (p_3)$  can be generically decomposed in terms of two form factors $F_{L,R}$ by
\begin{equation}
i {\cal M} = -i g \bar{u}_{\ell_k} (-p_2) (F_L P_L + F_R P_R) v_{\ell_m}(p_3) \, , 
\label{amplitude}
\end{equation}
and the partial decay width can then be written as follows:
\begin{eqnarray}
\Gamma (H \to {\ell_k} \bar{\ell}_m)& = &\frac{g^2}{16 \pi m_{H}} 
\sqrt{\left(1-\left(\frac{m_{\ell_k}+m_{\ell_m}}{m_{H}}\right)^2\right)
\left(1-\left(\frac{m_{\ell_k}-m_{\ell_m}}{m_{H}}\right)^2\right)} \nonumber\\
&& \times \Big((m_{H}^2-m_{\ell_k}^2-m_{\ell_m}^2)\big(|F_L|^2+|F_R|^2\big)- 4 m_{\ell_k} m_{\ell_m} Re(F_L F_R^{*})\Big) \, , 
\label{decay}
\end{eqnarray}
where $p_1$, $-p_2$ and $p_3$ are the ingoing Higgs boson momentum, the outgoing momentum of the lepton $\ell_k$ and the outgoing momentum of the antilepton $\bar \ell_m$, respectively, and having implemented the conservation of momentum by $p_1=p_3-p_2$. Besides, $m_H$ stands for the Higgs mass and $m_\ell=Y_\ell \, v$ for lepton masses (with $v=174$ GeV). The width of the $CP$-conjugate channel $H\to\ell_m \bar \ell_k$ is trivially related to the previous one  and their numerical values will coincide for the case of real Yukawa couplings.

We have performed a diagrammatic calculation of $\Gamma (H \to {\ell_k} \bar{\ell}_m)$ with the use of the MIA and considering the following points: (1) In contrast to the usual computations in the literature that work in the physical neutrino mass basis, we use instead the EW chiral neutrino basis; (2) However, we treat for convenience, the external particles $H$, $\ell_k$ and $\bar{\ell}_m$ in their physical mass basis; (3) We use the {\it fat propagator} for the heavy right-handed neutrinos and the Feynman rules as described in Section \ref{model}; (4) The LFVHD amplitude is evaluated at the one-loop order in the Feynman-'t Hooft gauge. In Appendix \ref{app:othergauge} it will be shown that the same result is obtained in the unitary gauge; (5) All the loops must contain one right-handed neutrino at least since they are the only particles transmitting LFV through the flavor off-diagonal neutrino Yukawa matrix entries; (6) According to the Feynman rules in figure~\ref{FR-MIA},  these flavor changing Yukawa couplings, $Y_{\nu}^{mk}$ with $m \neq k$, appear just in two places, the mass insertions given by $m_D$ and the interactions of the scalar sector with the left- and right-handed neutrinos being proportional to $Y_\nu$, therefore, the use of the MIA will provide a perturbative expansion in powers of $Y_\nu$; 7) Since $Y_{\nu}$ appears twice for each $\nu_R$ in an internal line, and because of the absence of interactions containing two right-handed neutrinos, all the one-loop diagrams will get an even number of powers of $Y_{\nu}$ (depending on the number of $\nu_R$'s). 

In summary, taking into account all the points exposed above, the one-loop contributions to the LFV Higgs decay amplitude, as computed with the MIA, will then be given by an expansion in even powers of $Y_\nu$. Concretely, with ${\cal O} (Y_\nu Y_\nu^{\dagger})$ being the leading order (LO) terms, 
${\cal O} (Y_\nu Y_\nu^{\dagger}Y_\nu Y_\nu^{\dagger})$ the next to leading order (NLO) terms, etc. Here, we consider the two most relevant contributions in this expansion, which in terms of the form factors of eq.~(\ref{amplitude}) can be written in the following way:
\begin{equation}
F_{L,R}^{{\rm MIA}\;\; (Y^2+Y^4)} = \left(Y_{\nu} Y_{\nu}^{\dagger} \right)^{km}f_{L,R}^{(Y^{2})} + \left(Y_{\nu} Y_{\nu}^{\dagger} Y_{\nu} Y_{\nu}^{\dagger}\right)^{km}f_{L,R}^{(Y^{4})} \, .
\label{FFMIA}
\end{equation}
We expect that, in the perturbativity regime of the neutrino Yukawa couplings, the next terms in this expansion, {\it i.e.} those of ${\cal O}(Y_{\nu}^{6})$ and the higher order terms, will be 
very tiny and can be safely neglected.  Furthermore, as will be explained in more detail in the next section, considering this expansion in powers of $Y_{\nu}$ and working with the hypothesis of $M_R$ being the heaviest scale, also lead to an implicit ordering of the various contributions in powers of $v/M_R$. In fact, we will demonstrate in the next section, by an explicit analytical expansion of the form factors in the large  $M_R\gg v $ limit, that the dominant terms of the two contributions in eq.(\ref{FFMIA}), the LO $f_{L,R}^{(Y^{2})}$ and the NLO $f_{L,R}^{(Y^{4})}$, indeed scale both as $(v/M_R)^2$. In contrast, the next order contributions, {\it i.e} those of ${\cal O}(Y_{\nu}^{6})$, scale as $(v/M_R)^4$, and therefore they will be negligible for heavy right-handed neutrinos, even when the Yukawa couplings are sizable. Therefore, considering just these two first terms in the MIA expansion, {\it i.e.} the LO and NLO terms of eq.(\ref{FFMIA}), will be sufficient to approach quite satisfactorily the full one-loop calculation of the neutrino mass basis in the case $\mu_X \ll m_D \ll M_R$ that we are interested in.
  
\begin{figure}[t!]
\begin{center}
\includegraphics[scale=0.8]{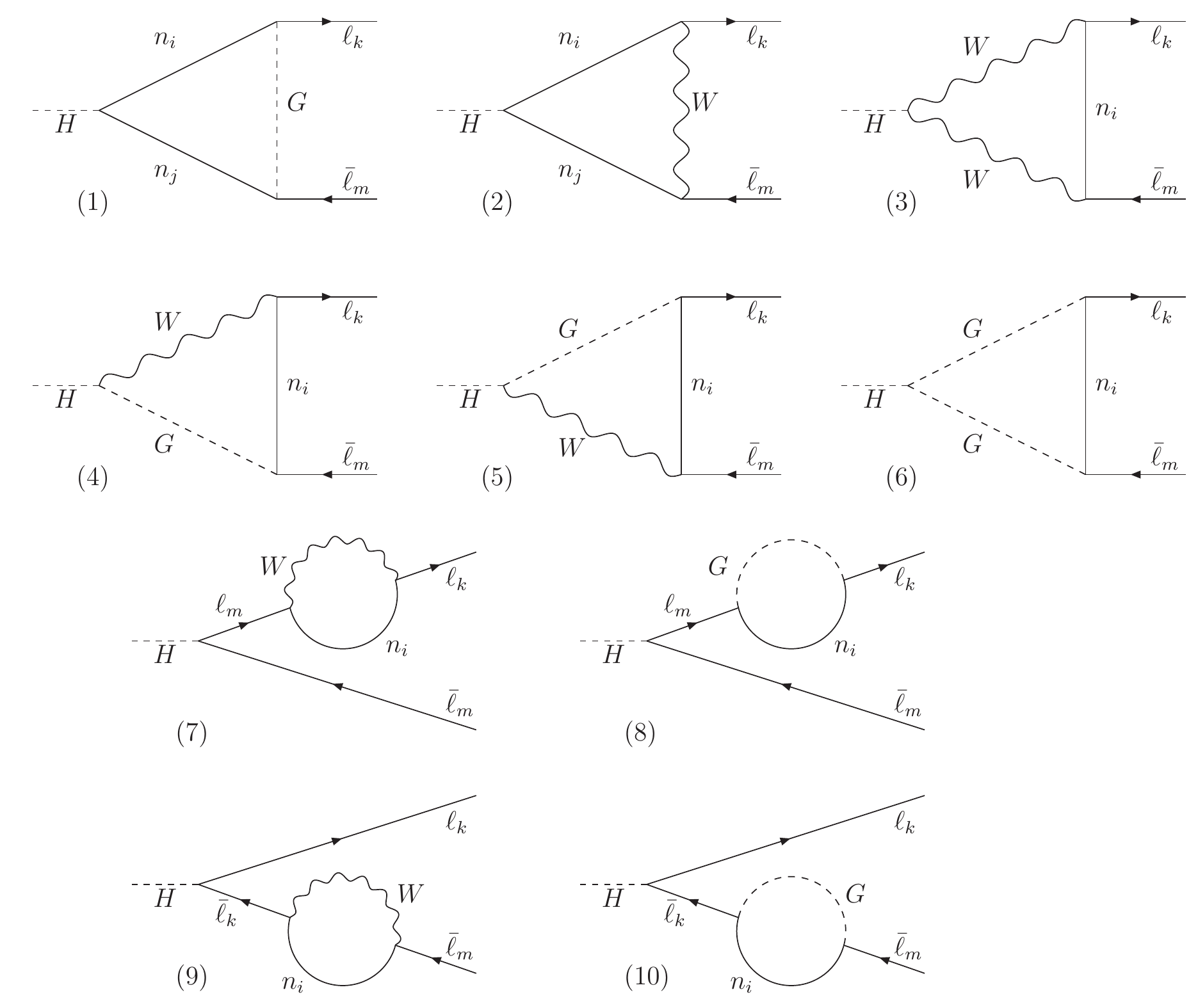}
\caption{One-loop diagrams  contributing to the full computation of $H\to\ell_k {\bar \ell_m}$ decays in the physical neutrino mass eigenstate basis.}
\label{DiagramsMassbasis}
\end{center}
\end{figure}

In order to estimate the goodness of the MIA in the present study of the LFV Higgs decays  we also include in this work a numerical comparison of our MIA results for the LFVHD rates with those of the full one-loop computation done in the physical particle mass basis which we take from \cite{Arganda:2004bz} and \cite{Arganda:2014dta}. For an easy comparison, we adopt in the MIA the same notation (i) (i=1,...,10) for the ten types of generic diagrams as in the full computation of \cite{Arganda:2004bz}. 
The full one-loop results will  then be computed in the Feynman-'t Hooft gauge as in that reference by adding the contributions of the 10 one-loop diagrams of the mass basis\footnote{We have noticed that with the sign conventions of the Feynman rules defined in our figure \ref{FR-MIA} our results for the contributions from diagrams (1), concretely $F_L^{(1)}$, (4) and (5) in the full one-loop computation get an opposite sign with respect to those in \cite{Arganda:2004bz}. However, we have checked that these detected typos do not affect the present comparison between the full one-loop computation and the MIA results.}, which for clarity are shown in figure \ref{DiagramsMassbasis}.  They classify into diagrams with vertex corrections, i=1,..,6,  and diagrams with external leg corrections, i=7,..,10.

\begin{figure}[t!]
\begin{center}
\includegraphics[scale=0.8]{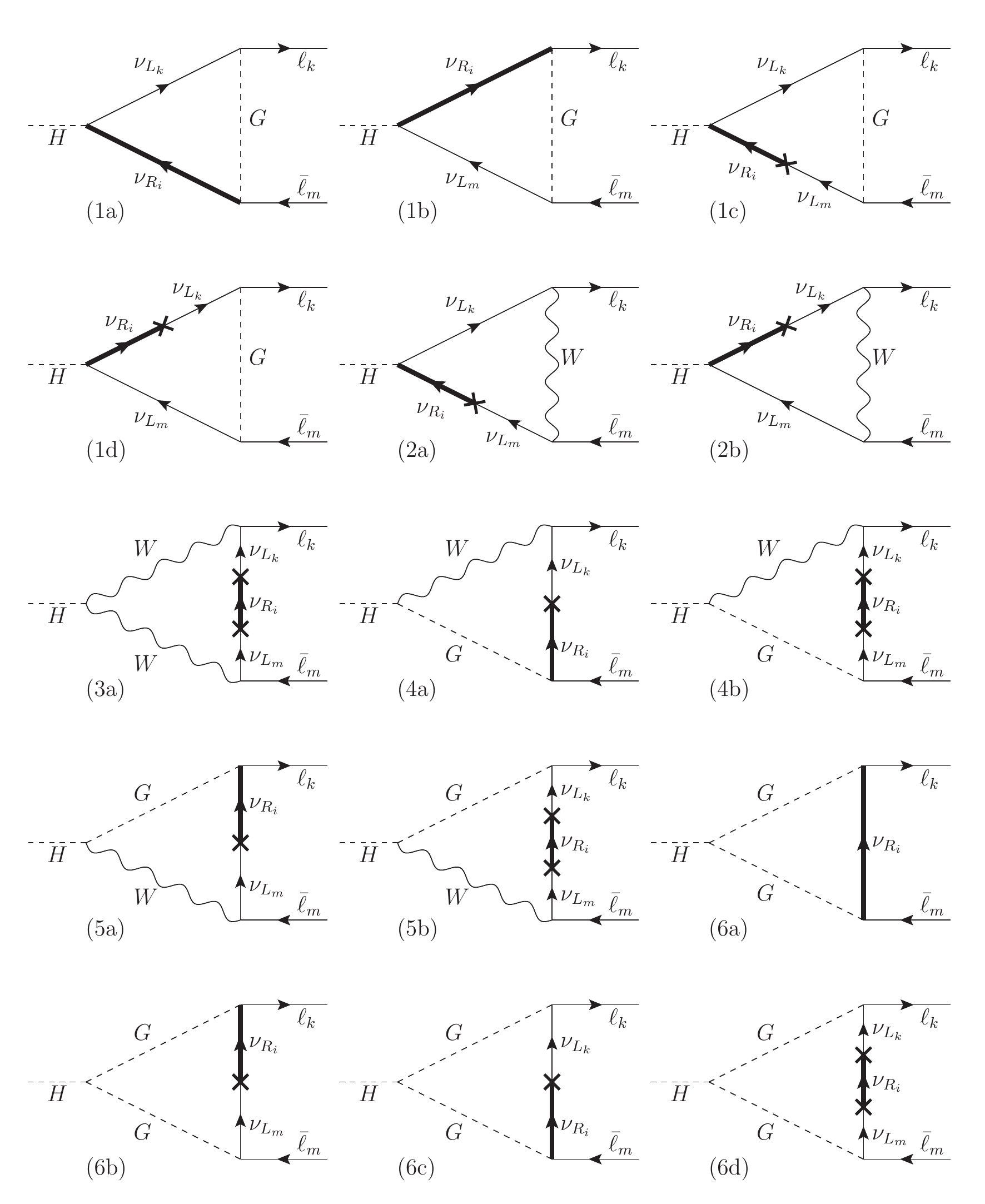}
\caption{Relevant vertex diagrams for the MIA form factors of LFVHD to ${\cal O}(Y_\nu^2)$.}
\label{diagsvertexY2}
\end{center}
\end{figure}

\begin{figure}[t!]
\begin{center}
\includegraphics[scale=0.8]{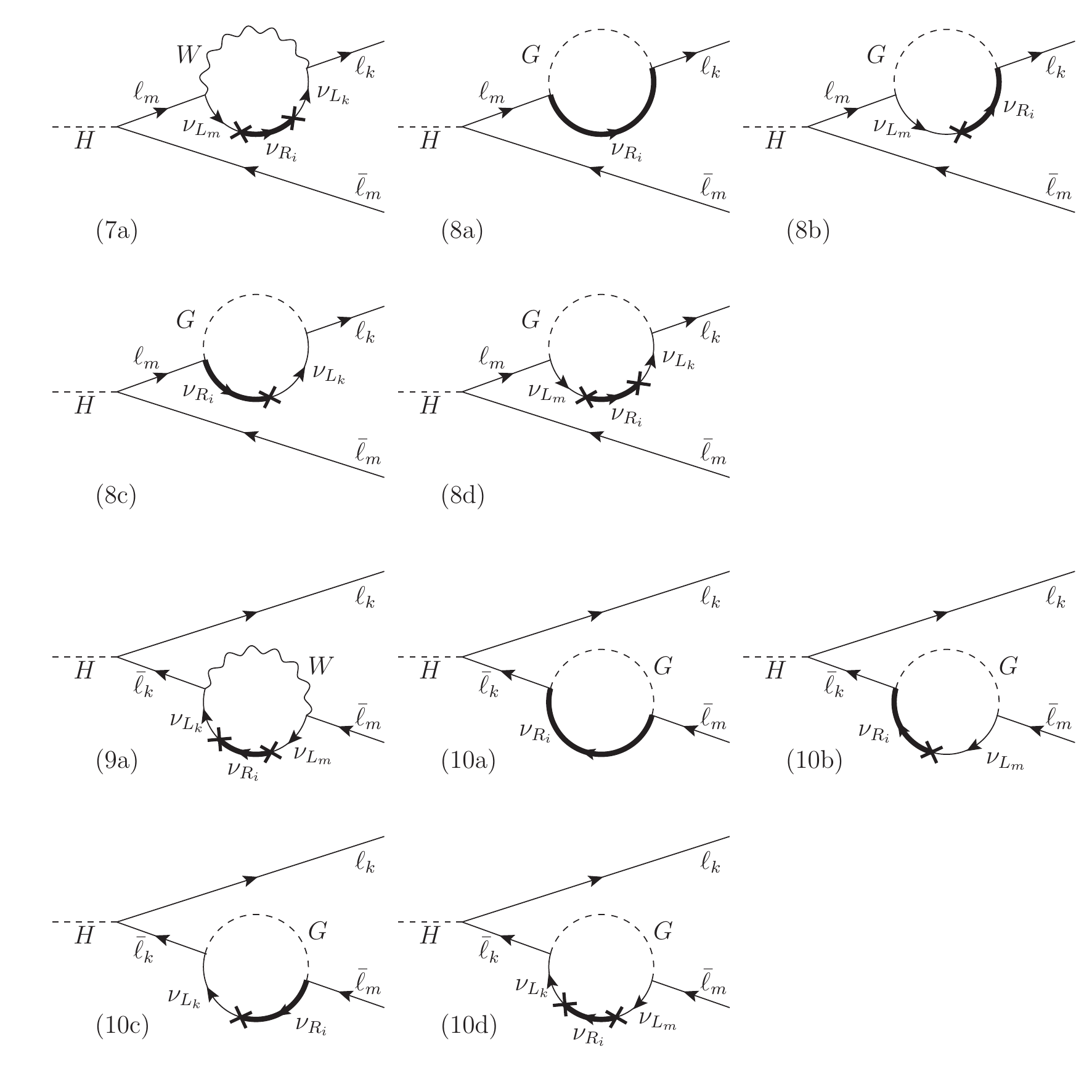}
\caption{Relevant external leg diagrams for the MIA form factors of LFVHD to ${\cal O}(Y_\nu^2)$.}
\label{diagslegY2}
\end{center}
\end{figure}

\begin{figure}[t!]
\begin{center}
\includegraphics[scale=0.8]{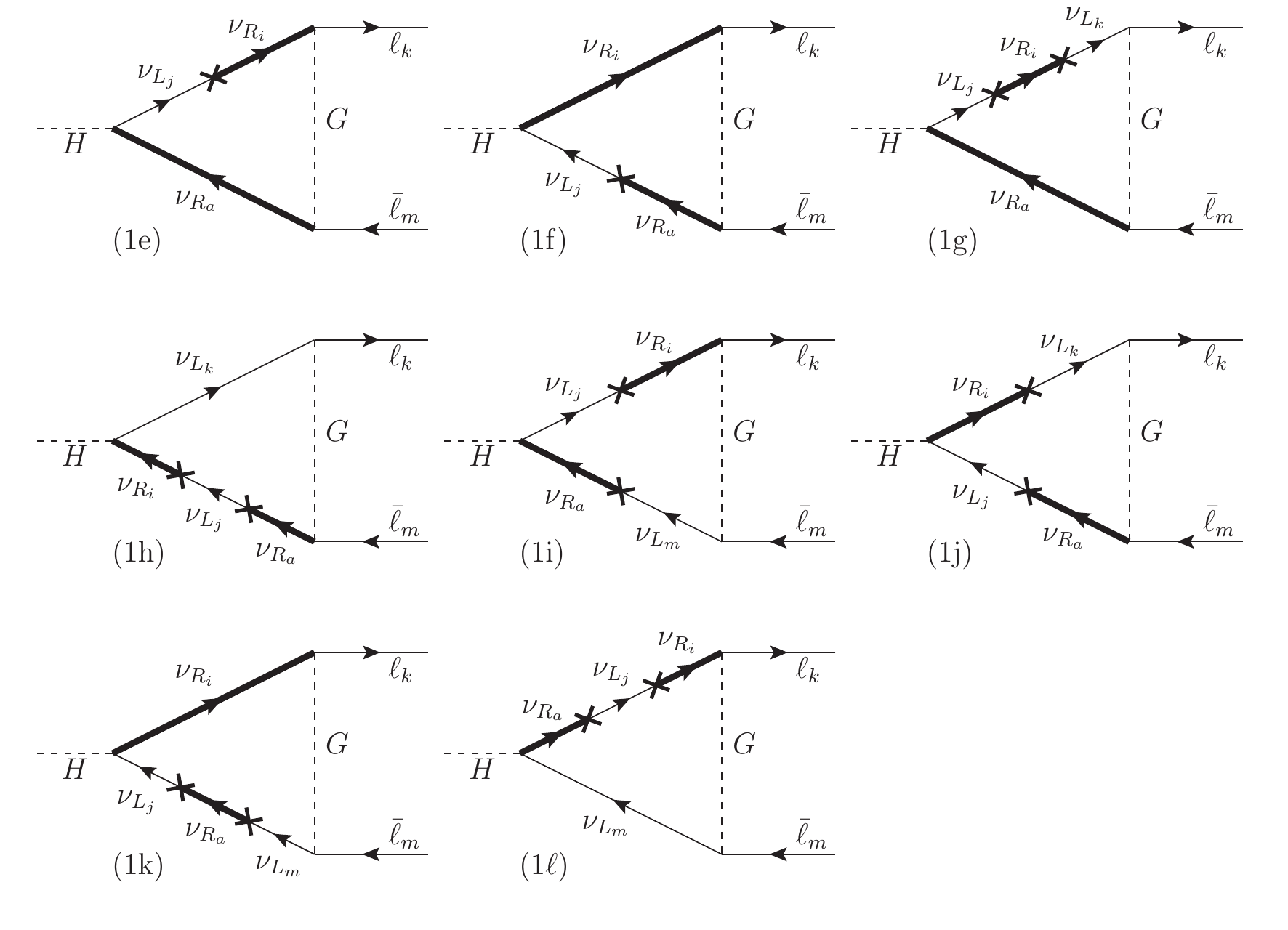}
\caption{Relevant vertex diagrams for the MIA form factors of LFVHD to ${\cal O}(Y_\nu^4)$.}
\label{diagsvertexY4}
\end{center}
\end{figure}

\begin{figure}[t]
\begin{center}
\includegraphics[scale=0.8]{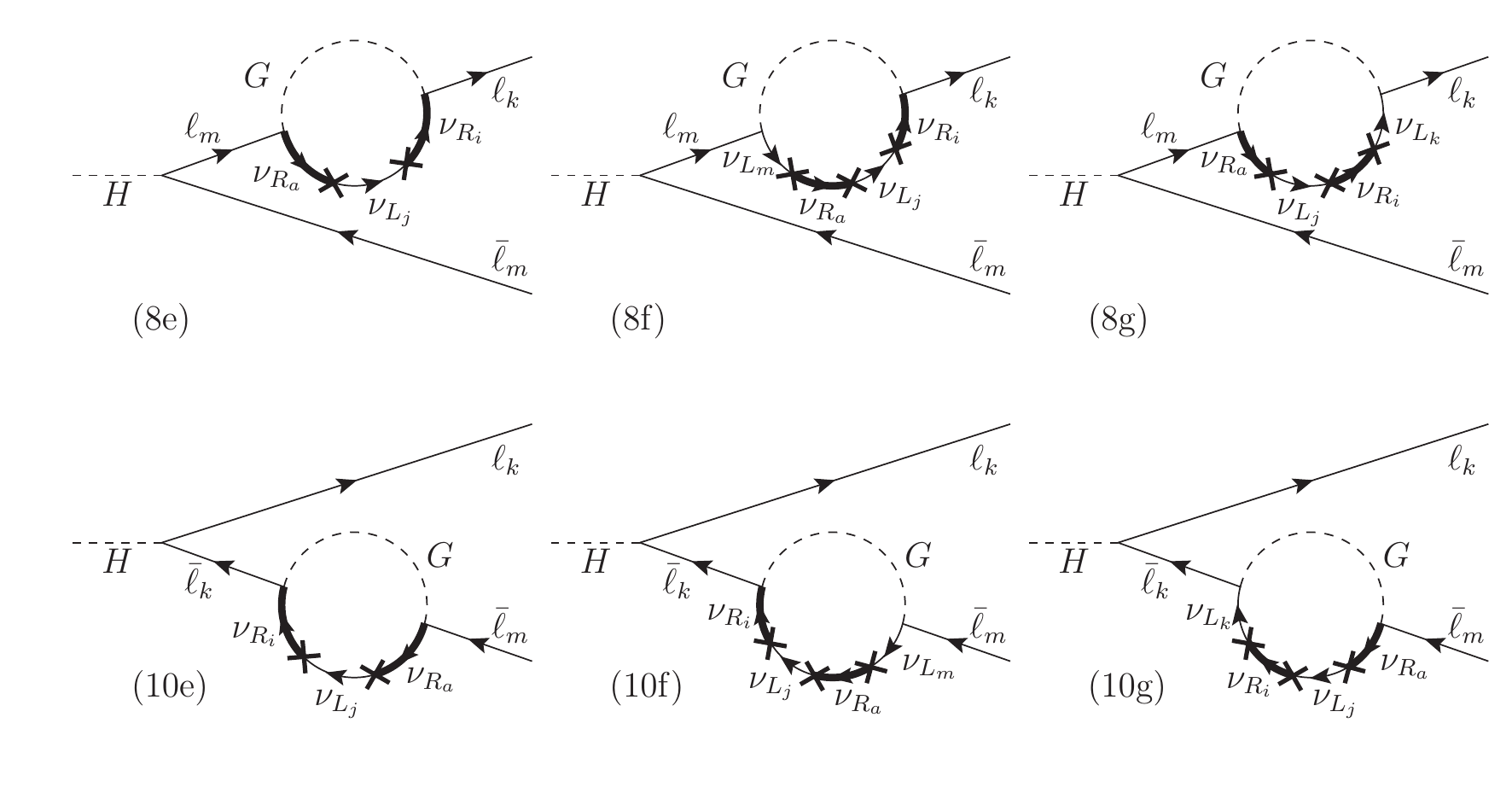}
\caption{Relevant external leg diagrams for the MIA form factors of LFVHD to ${\cal O}(Y_\nu^4)$.}
\label{diagslegY4}
\end{center}
\end{figure}

For the one-loop computation in the MIA, we also follow a diagrammatic procedure that consists of the systematic insertion of right-handed  neutrino (fat) propagators in all the possible places inside the loops which are built with the interaction vertices and propagators of figure~\ref{FR-MIA}. Generically, diagrams with one  right-handed neutrino propagator will contribute to the form factors of ${\cal O}(Y_{\nu}^{2})$, whereas diagrams with two right-handed neutrino propagators will contribute to the form factors of ${\cal O}(Y_{\nu}^{4})$.   In figures \ref{diagsvertexY2}, \ref{diagslegY2}, \ref{diagsvertexY4} and \ref{diagslegY4}, we show the relevant one-loop diagrams in the MIA corresponding to the dominant contributions of the LO and the NLO respectively. These are also classified into those of vertex corrections type and those of leg corrections type. 
The MIA form factors are then obtained accordingly as the sum of all these contributions that can be summarized as follows:
\begin{equation}
F_{L,R}^{{\rm MIA}}=\sum_{{\rm i}=1}^{10} F_{L,R}^{\rm MIA (i)}\,.
\label{FFLRMIA}
\end{equation}
At ${\cal O}(Y_{\nu}^{2})$,  each  $F_{L,R}^{\rm MIA (i)}$ receives contributions from diagrams all containing 1 right-handed neutrino propagator and one of these three combinations: (i) 1 vertex with $\nu_R$ and 1 $m_D$ insertion,  
(ii)  0 vertices with $\nu_R$ and 2 $m_D$ insertions, (iii)  2 vertices with $\nu_R$ and 0 $m_D$ insertions. 
This drives to the relevant  diagrams in figures  \ref{diagsvertexY2}, \ref{diagslegY2} whose contributions are given, in an obvious  correlated notation, by:
\begin{eqnarray}
F_{L,R}^{{\rm MIA (1)\,\, (Y^2)}}&=&F_{L,R}^{\rm  (1a)}+F_{L,R}^{\rm  (1b)}+F_{L,R}^{\rm  (1c)}+F_{L,R}^{\rm  (1d)}\,, \nonumber \\  
F_{L,R}^{{\rm MIA (2)\,\, (Y^2)}}&=&F_{L,R}^{\rm  (2a)}+F_{L,R}^{\rm  (2b)} \,,
\nonumber \\  
F_{L,R}^{{\rm MIA (3)\,\, (Y^2)}}&=&F_{L,R}^{\rm  (3a)} \,,
\nonumber \\ 
F_{L,R}^{{\rm MIA (4)\,\, (Y^2)}}&=&F_{L,R}^{\rm  (4a)}+F_{L,R}^{\rm  (4b)} \,,
\nonumber \\ 
F_{L,R}^{{\rm MIA (5)\,\, (Y^2)}}&=&F_{L,R}^{\rm  (5a)}+F_{L,R}^{\rm  (5b)} \,,
\nonumber \\ 
F_{L,R}^{{\rm MIA (6)\,\, (Y^2)}}&=&F_{L,R}^{\rm  (6a)}+F_{L,R}^{\rm  (6b)}+F_{L,R}^{\rm  (6c)}+F_{L,R}^{\rm  (6d)} \,,
\nonumber \\ 
F_{L,R}^{{\rm MIA (7)\,\, (Y^2)}}&=&F_{L,R}^{\rm  (7a)} \,,
\nonumber \\ 
F_{L,R}^{{\rm MIA (8)\,\, (Y^2)}}&=&F_{L,R}^{\rm  (8a)}+F_{L,R}^{\rm  (8b)}+F_{L,R}^{\rm  (8c)}+F_{L,R}^{\rm  (8d)} \,,\nonumber \\  
F_{L,R}^{{\rm MIA (9)\,\, (Y^2)}}&=&F_{L,R}^{\rm  (9a)} \,,
\nonumber \\ 
F_{L,R}^{{\rm MIA (10)\,\, (Y^2)}}&=&F_{L,R}^{\rm  (10a)}+F_{L,R}^{\rm  (10b)}+F_{L,R}^{\rm  (10c)}+F_{L,R}^{\rm  (10d)} . 
\label{FFLRMIAY2}
\end{eqnarray}
At ${\cal O}(Y_{\nu}^{4})$,  each  $F_{L,R}^{\rm MIA (i)}$ receives contributions from diagrams all containing 2 right-handed neutrino propagators and one of these three combinations: i) 2 vertices with $\nu_R$ and 2 $m_D$ insertions,  
ii)  3 vertices with $\nu_R$ and 1 $m_D$ insertion, iii)  1 vertex with $\nu_R$ and 3 $m_D$ insertions. Other possible combinations will provide subleading corrections in the heavy $M_R$ case of our interest,  since they will come with extra powers of $M_R$ in the denominator. Thus, we find that the most relevant diagrams are those of type (1), (8) and (10) summarized in  figs. \ref{diagsvertexY4} and \ref{diagslegY4}, whose respective contributions are given by:
\begin{eqnarray}
F_{L,R}^{{\rm MIA (1)\,\, (Y^4)}}&=&F_{L,R}^{\rm  (1e)}+F_{L,R}^{\rm  (1f)}+F_{L,R}^{\rm  (1g)}+F_{L,R}^{\rm  (1h)} 
+F_{L,R}^{\rm  (1i)}+F_{L,R}^{\rm  (1j)}+F_{L,R}^{\rm  (1k)}+F_{L,R}^{\rm  (1\ell)} \,,
\nonumber \\  
F_{L,R}^{{\rm MIA (8)\,\, (Y^4)}}&=&F_{L,R}^{\rm  (8e)}+F_{L,R}^{\rm  (8f)}+F_{L,R}^{\rm  (8g)}  \,,
\nonumber \\  
F_{L,R}^{{\rm MIA (10)\,\, (Y^4)}}&=&F_{L,R}^{\rm  (10e)}+F_{L,R}^{\rm  (10f)}+F_{L,R}^{\rm  (10g)}  .
\label{FFLRMIAY4}
\end{eqnarray}
The explicit analytical results for all the form factors above, $F_{L,R}^{{\rm MIA (i)\,\, (Y^2)}}$ with i=1,..10, and $F_{L,R}^{{\rm MIA (i)\,\, (Y^4)}}$ with i=1,8,10, are given in eqs.~(\ref{FLtot_op2_Y2})-(\ref{FRtot_op2_Y4}) of the Appendix \ref{app:FormFactors}. These results are expressed in terms of the usual  one-loop Veltman-Passarino functions of two points ($B_{0}$ and $B_{1}$), three points ($C_0$, $C_{11}$, $C_{12}$ and $\tilde{C}_{0}$) and four points ($D_{12}$, $D_{13}$ and $\tilde{D}_{0}$) whose definitions are given in eqs.~(\ref{loopfunctionB})-(\ref{loopfunctionD}).

 Some comments about the analytical properties of the previous MIA results are in order. First, we analyze their  ultraviolet behavior.  From \cite{Arganda:2004bz}, we know that in the full one-loop computation of the mass basis only the contributions to the amplitude from diagrams (1),  (8) and (10) of figure \ref{DiagramsMassbasis} are ultraviolet divergent separately, and the total sum from these diagrams (1)+(8)+(10) is finite, therefore providing a total one-loop amplitude that is ultraviolet finite as it must be. We have checked again this same result of the full computation. In addition we have explored the divergences of the MIA diagrams. Our calculation in the MIA also shows that diagrams of type (2), (3), (4), (5), (6), (7) and (9) are convergent separately, while each contribution of ${\cal O}(Y_{\nu}^{2})$ from diagrams (1), (8) and (10) are divergent but their divergences cancel out again in their sum. For this reason, in the next numerical analysis, whenever we present results for each diagram we will consider  the sum (1)+(8)+(10), which is convergent and therefore meaningful,  instead of the contributions from each of these diagrams separately. 
 
 Second, it is also worth to comment on the gauge invariance of our previous MIA results for the decay amplitude. Remember that our computation in this section has been performed in the Feynman-'t Hooft gauge. In order to prove the gauge invariance of our results, we have computed the amplitude also in other gauges and checked that we get the same result. Specifically,  we have computed the form factors $F_{L,R}$  in the unitary gauge and in arbitrary $R_{\xi}$ gauges. The details of the unitary gauge computation are collected in Appendix \ref{app:othergauge}. 
 
 Finally, we present our numerical analysis. 
For the forthcoming numerical comparisons of the full and the MIA results, and in order to find out the goodness of this approximation we need to fix the numerical values of the input model parameters. Particularly relevant are the input $Y_\nu$ coupling matrix and the input $M_R$. Regarding the heavy mass $M_R$ we will explore numerical values in a wide range, say 
between 200 GeV and 15 TeV. For $Y_\nu$ we will select several illustrative examples which have been used in the literature.  However, we wish to emphasize that our analytical results could be applied to other input values, and this is in fact our final purpose, since we believe that our formulas may be useful for the community who wishes to perform  an easy and  fast numerical estimate  of the LFVHD rates with their own choices for the input parameters $Y_\nu$, and $M_R$,  and without the need to rotate to the physical basis which, depending on the examples, may involve a heavy numerical work. 

We choose the following five examples of $Y_\nu$ for the numerical estimates and for the comparisons of the full versus the MIA results. The first four examples are taken from \cite{Arganda:2014dta, Arganda:2015ija, Arganda:2015naa,DeRomeri:2016gum}, where  these particular textures together with others were selected as belonging to a type of scenarios in which the LFV is always extremely suppressed in the $\mu e$ sector but it can lead to large LFV in either the $\tau \mu$ (named TM scenarios) or in the $\tau e$ sectors (named TE scenarios), although never producing LFV in these two sectors simultaneously. These scenarios are known to produce interesting phenomenological implications in collider physics as, for instance, the production of asymmetric $\tau$-$\mu$-jet-jet and $\tau$-$e$-jet-jet events at LHC~\cite{Arganda:2015ija}. 
The Yukawa coupling matrices in these examples are usually given in terms of a  scaling factor $f$ that characterizes the global strength of the coupling. We consider in particular the following four examples:
\begin{align}
Y_{\nu}^{\rm TM4}=f\left(\begin{array}{ccc}
0.1&0&0\\0&1&0\\0&1&0.014
\end{array}\right)\,, &\quad
Y_{\nu}^{\rm TM5}=f\left(\begin{array}{ccc}
0&1&-1\\0.9&1&1\\1&1&1
\end{array}\right) \,,\\ 
Y_{\nu}^{\rm TM9}=f\left(\begin{array}{ccc}
0.1&0&0\\0&0.46&0.04\\0&1&1
\end{array}\right)\,, &\quad
Y_{\nu}^{\rm TE10}=f\left(\begin{array}{ccc}
0.94&0&0.08\\0&0.1&0\\1&0&-1
\end{array}\right)\,. 
\label{YnuTMTE}
\end{align}
 
We introduce next the last example, called here GF, which we have deduced from the results in \cite{Fernandez-Martinez:2016lgt}. In this and other references (see, for instance \cite{Antusch:2006vwa,FernandezMartinez:2007ms,delAguila:2008pw,Antusch:2008tz,Antusch:2014woa,Fernandez-Martinez:2015hxa}), there have been explored the constraints that are imposed on the Yukawa matrix entries by means of global fits (GF) to data. More concretely, the constraints are set into the product $Y_\nu Y_\nu^\dagger$ by means of another matrix $\eta$ that is related to the Yukawa matrix approximately by: $\eta = (v^2/(2M_R^2))(Y_\nu Y_\nu^\dagger)$. Then, we choose our third example $Y_\nu^{\rm GF}$ such that it saturates, at the $3\sigma$ level,  the present experimental constraints given in \cite{Fernandez-Martinez:2016lgt}. More concretely, these constraints define a maximum allowed by data $\eta$ matrix given by:
\begin{align}
 \eta_{3\sigma}^{\rm max}=\left(\begin{array}{ccc}
1.62\times 10^{-3}&1.51\times 10^{-5}&1.57\times 10^{-3}\\1.51\times 10^{-5}&3.92\times 10^{-4}&9.24\times 10^{-4}
\\1.57\times 10^{-4}&9.24\times 10^{-4}&3.67\times 10^{-3}
\end{array}\right)\,,
\label{etamax3sigma}
\end{align}
and this can be reached, for instance,  by our choice: 
\begin{align}
 Y_{\nu}^{\rm GF}=f\left(\begin{array}{ccc}
 0.33&0.83&0.6\\-0.5&0.13&0.1\\-0.87&1&1
 \end{array}\right)\,,
 \label{YnuGF}
\end{align}
in a parameter space line given by the ratio $f/M_R= (3/10)\, {\rm TeV}^{-1}$, i.e. for 
$(f,M_R)=(3,10\, {\rm TeV})$,  $(1,3.3\, {\rm TeV})$,  $(0.3,1\, {\rm TeV})$, \dots, etc.
We have checked that other choices for $Y_{\nu}$, like those for the other TM and TE  scenarios defined in
\cite{Arganda:2014dta, Arganda:2015ija, Arganda:2015naa,DeRomeri:2016gum}, and also the $Y_{\nu}$ taken in \cite{Abada:2014kba},  lead to similar conclusions regarding the goodness of the MIA approximation as those ones that we study here.

\subsection{Goodness of the MIA results to $\boldsymbol{{\cal O}(Y_{\nu}^{2})}$}
\begin{figure}[t!]
\begin{flushright}
\includegraphics[width=.95\textwidth]{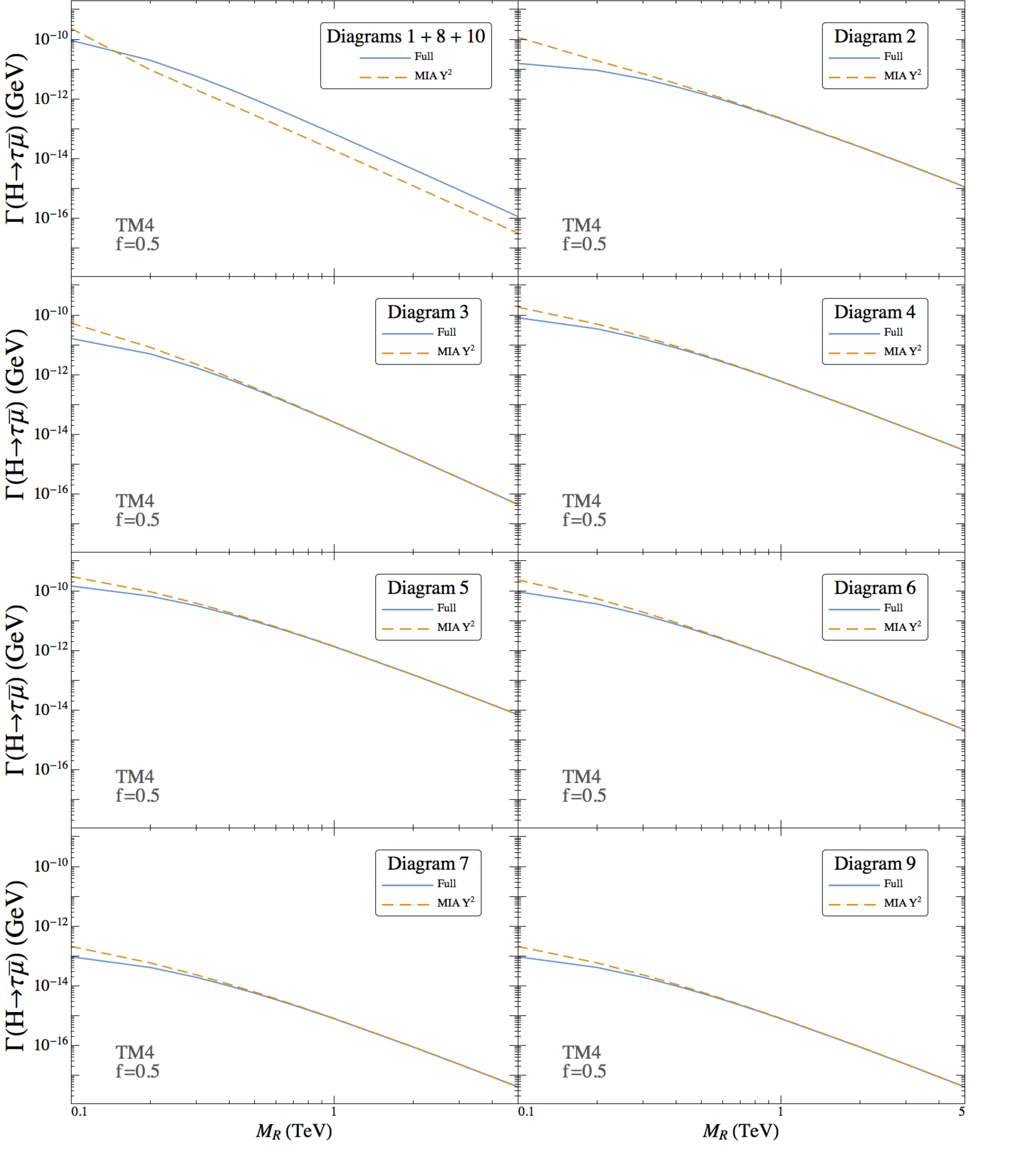}
\caption{Predictions for the contributions from the various diagrams to the partial width $\Gamma (H \to \tau \bar \mu)$ as a function of $M_R$.  The dashed lines are the predictions from the MIA to ${\cal O}(Y_\nu^2)$. The solid lines are the predictions from the full-one loop computation of the mass basis. Here the example TM4 with f=0.5, as explained in the text,  is chosen.}
\label{diagbydiagcomparisonY2}
\end{flushright}
\end{figure}
We start with the first order results in the mass insertion approximation. Therefore for the numerical evaluation we use our formulas for the form factors to ${\cal O}(Y_{\nu}^{2})$, given in eqs.~(\ref{FLtot_op2_Y2}) and (\ref{FRtot_op2_Y2}) of the Appendix \ref{app:FormFactors}. We first show in figure \ref{diagbydiagcomparisonY2} the partial decay width of the full calculation together with our predictions from the MIA to ${\cal O}(Y_{\nu}^{2})$, and we have separated explicitly the results from the various (i) diagrams [recall that we have considered the sum of diagrams (1), (8) and (10) in order to have a finite contribution]. We will explore in these plots the comparative predictions of the full versus the MIA computations as a function of $M_R$ in order to conclude on the goodness of this approximation with respect to this input parameter mass. We will discuss the particular case of the scale factor $f=0.5$ of the neutrino Yukawa coupling as a reference value.  

As we see from figure \ref{diagbydiagcomparisonY2}, the contribution from each diagram [or group of diagrams in the case of (1)+(8)+(10)] to the form factor and in consequence to the width decreases with $M_R$. In fact, this behavior can be very well understood with our simple formulas of the large $M_R$ expansions in eq.~(\ref{FLsimple_totdom}). In particular, when adding the three contributions (1)+(8)+(10) in the MIA we see explicitly the cancellation of the divergent contributions from $\Delta$ terms and the corresponding cancellation of the regularization $\mu$ scale dependent terms. The final behavior of the remaining finite terms in each form factor go generically as $\sim (v^2/M_R^2)$, and in addition there are also logarithmic terms going as $\sim (v^2/M_R^2) ({\rm Log}(v^2/M_R^2))$. 

\begin{figure}[t!]
\begin{center}
\includegraphics[width=.49\textwidth]{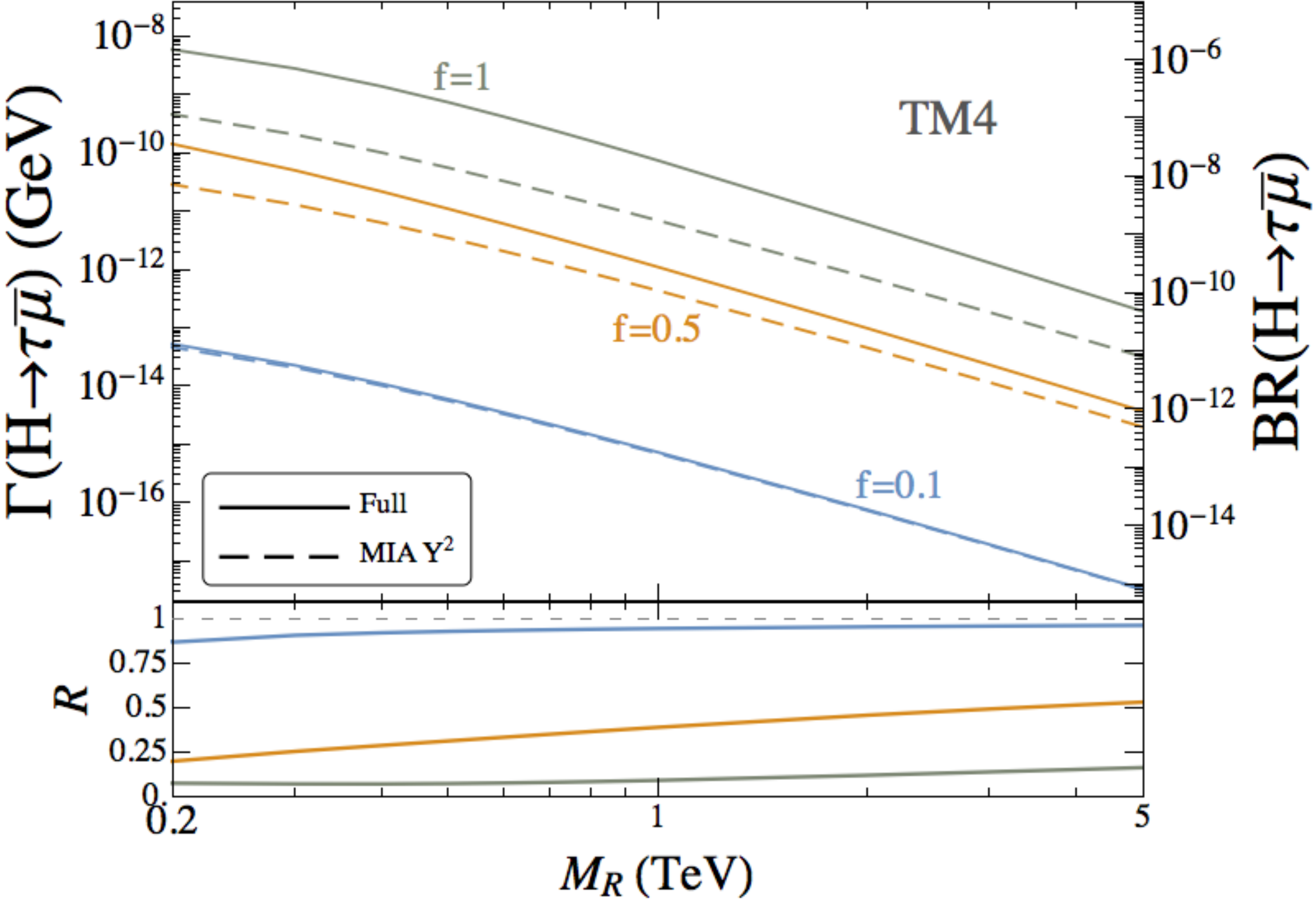}
\includegraphics[width=.49\textwidth]{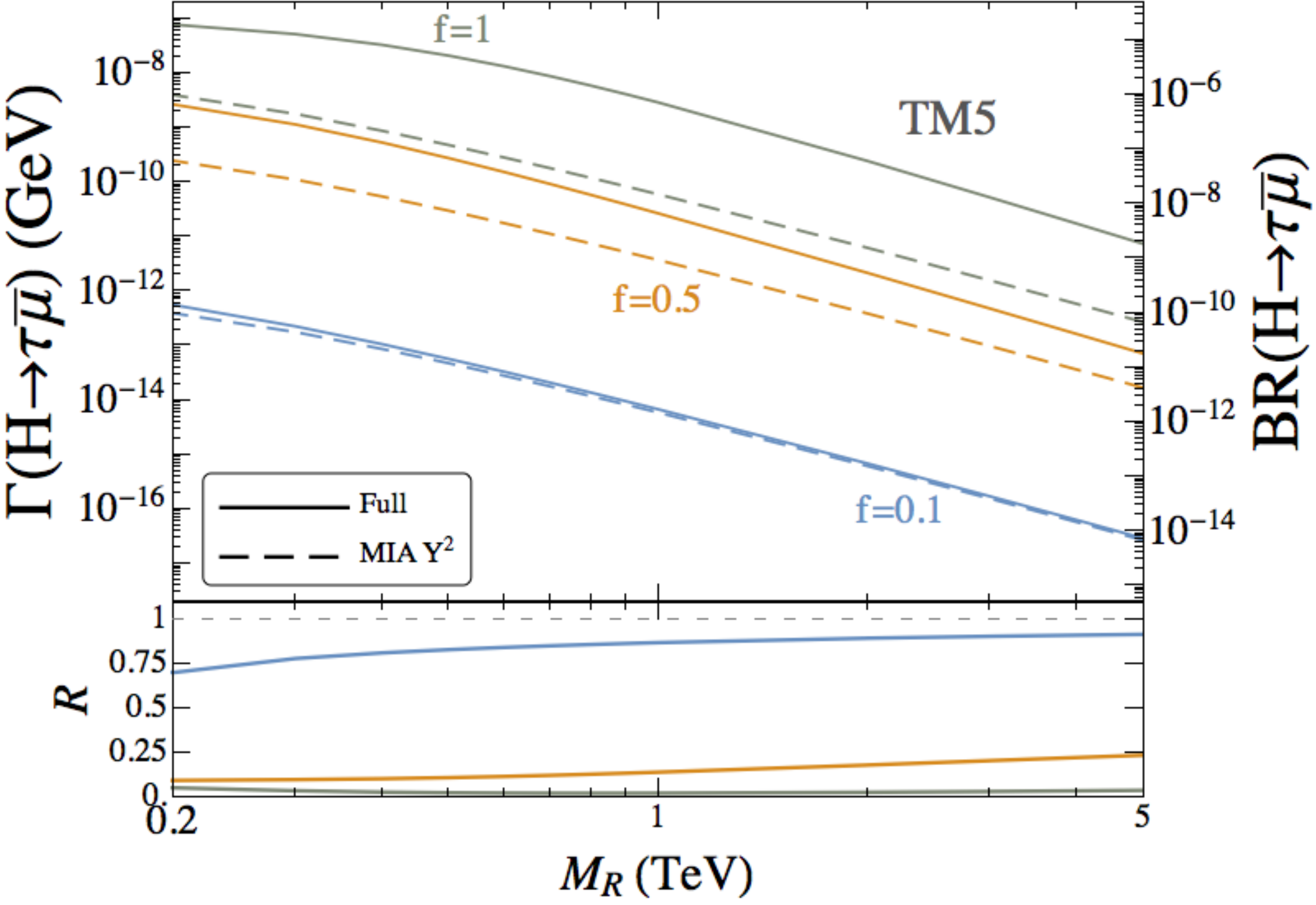}
\caption{Predictions for the partial width $\Gamma (H \to \tau \bar \mu)$ and branching ratio BR$(H \to \tau \bar \mu)$ as a function of $M_R$.  The dashed lines are the predictions from the MIA to ${\cal O}(Y_\nu^2)$. The solid lines are the predictions from the full-one loop computation of the mass basis. Here the examples TM4 (left panel) and TM5 (right panel) with f=0.1,0.5,1, as explained in the text,  are chosen. In the bottom of these plots the ratio $R=\Gamma_{\rm MIA}/ \Gamma_{\rm full}$ is also shown. }\label{totalcomparisonY2}
\end{center}
\end{figure}

We observe a consistent agreement between the MIA and full results for diagrams (2), (3), (4), (5), (6), (7) and (9). For the sum (1)+(8)+(10), the MIA reproduces the behavior of the full calculation very well but there is a mismatch in this example by an approximate factor of 3 in the partial decay width. In figure \ref{totalcomparisonY2}, we observe that this difference translates to the total sum. However, if we decrease the value of $f$, both calculations are in very good agreement. On the other hand, the larger is $f$, the worse is the discrepancy between them. In order to give a quantitative statement on this observation, we define the ratio $R=\Gamma_{\rm MIA}/ \Gamma_{\rm full}$. From the bottom of figure \ref{totalcomparisonY2}, we have $R$ close to 1 for low values of $f$ ($f=0.1$) and large $M_R$ above 1 TeV. If we increase $f$ up to 1, poor values of $R$ far from 1 are obtained in the full $M_R$ interval studied, so the MIA results to ${\cal O}(Y_{\nu}^{2})$ do not reproduce satisfactorily the full calculation results. For this particular set of parameters, the branching ratios are still very far from the scope of the LHC Run II, thus at this level of approximation, large values of $f$ are interesting from a phenomenological point of view.

We conclude that for large values of $f$, we need to include the next to leading order terms in the MIA expansion.  In consequence, for small values of $Y_{\nu}$, the MIA results up to ${\cal O}(Y_{\nu}^{2})$ are in very good agreement with the full results. For large values of $Y_{\nu}$, the MIA only reproduces the functional behavior but not the numerical values. Thus, it is necessary to include in the MIA computation the next order contributions,  i.e. ${\cal O}(Y_{\nu}^{4})$.

\subsection{Goodness of the  MIA results to $\boldsymbol{{\cal O}(Y_{\nu}^{2}+Y_{\nu}^{4})}$}

\begin{figure}[t!]
\begin{center}
\includegraphics[width=.49\textwidth]{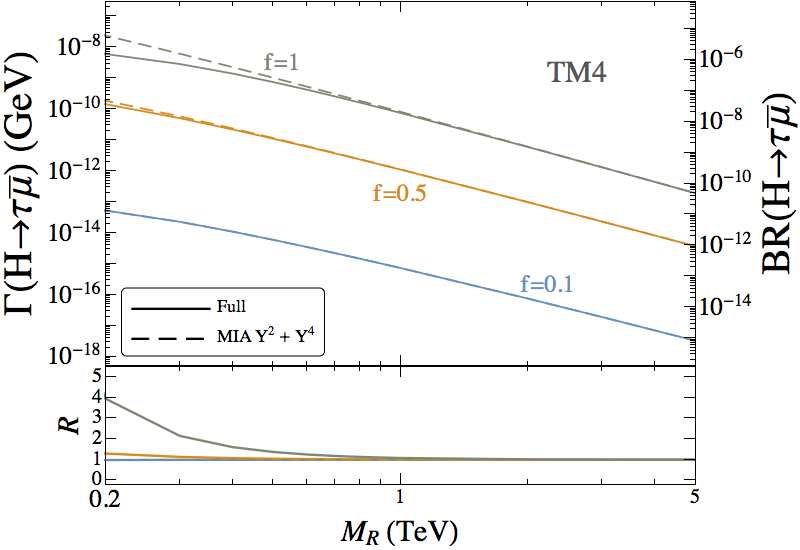}
\includegraphics[width=.49\textwidth]{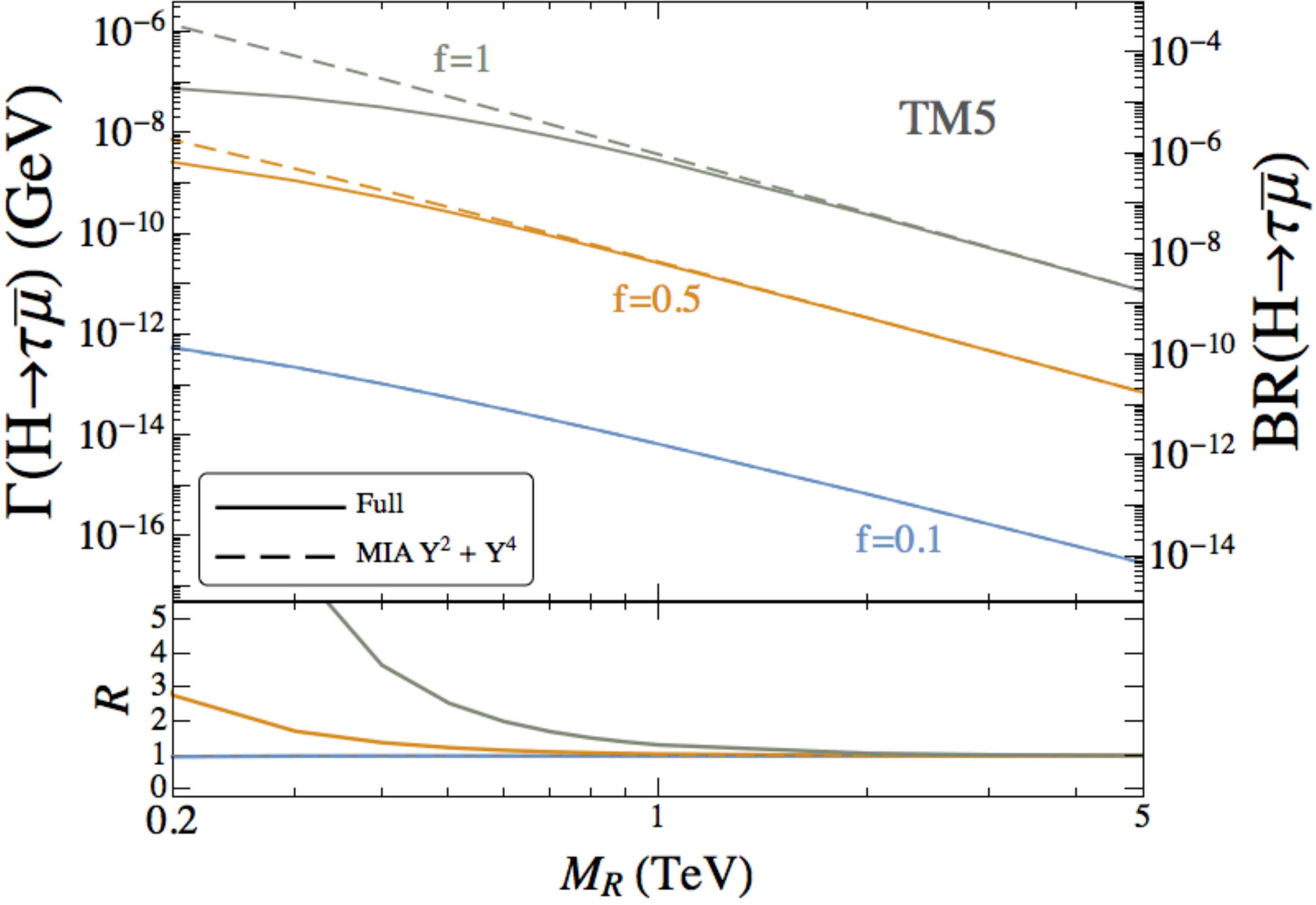}
\caption{Predictions for the partial width $\Gamma (H \to \tau \bar \mu)$ and branching ratio BR$(H \to \tau \bar \mu)$ as a function of $M_R$.  The dashed lines are the predictions from the MIA to ${\cal O}(Y_\nu^2+Y_\nu^4)$. The solid lines are the predictions from the full-one loop computation of the mass basis. Here the examples TM4 (left panel) and TM5 (right panel) with f=0.1,0.5,1, as explained in the text,  are chosen. In the bottom of these plots the ratio $R=\Gamma_{\rm MIA}/ \Gamma_{\rm full}$ is also shown. }
\label{totalcomparisonY2Y4}
\end{center}
\end{figure}

As we learn from of eqs.~(\ref{FLtot_op2_Y4}) and (\ref{FRtot_op2_Y4}) in the Appendix \ref{app:FormFactors}, the dominant contributions of ${\cal O}(Y_{\nu}^{4})$ come from diagrams (1), (8) and (10). We have seen in figure \ref{diagbydiagcomparisonY2} that the other diagrams are well described by the ${\cal O}(Y_{\nu}^{2})$ terms. After including all the relevant  ${\cal O}(Y_{\nu}^{2}+Y_{\nu}^{4})$ contributions, we see in figure \ref{totalcomparisonY2Y4} that the total sum of MIA diagrams is in very good agreement with the full results for different values of $f$.

Therefore, we can conclude that the MIA calculation, with the inclusion of the most relevant ${\cal O}(Y_{\nu}^{4})$ terms, corrects the ${\cal O}(Y_{\nu}^{2})$ contributions and achieves a better fit to the full numerical results for this process in the large $M_{R}\gg v Y_\nu$ mass range.  In particular, we see this improvement with respect to ${\cal O}(Y_{\nu}^{2})$ contributions from the closeness of $R$ to 1 for different values of $f$. 
How large $M_{R}$ should be in order to get a good numerical prediction of the LFVHD rates  depends obviously on the size of the Yukawa coupling.  For small Yukawa coupling, i.e. for small $f\lesssim 0.5$ the MIA works pretty well for  $M_R$ above 400 GeV, whereas for larger couplings, say $f$ above 0.5,  the MIA also provides a good result but requires heavier $M_R$, above 1000 GeV. 
 
\begin{figure}[t!]
\begin{center}
\includegraphics[width=.49\textwidth]{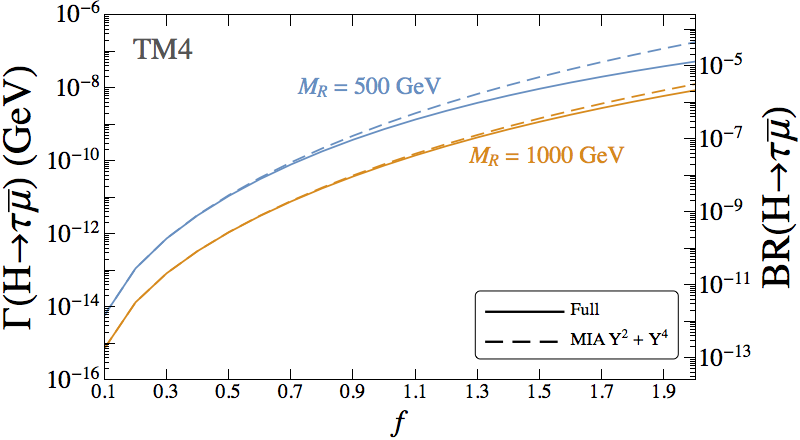}
\includegraphics[width=.49\textwidth]{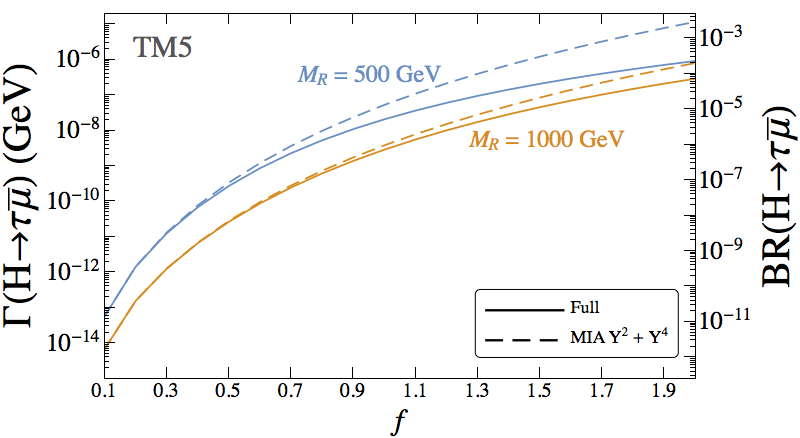}
\caption{Predictions for the partial width $\Gamma (H \to \tau \bar \mu)$ and branching ratio BR$(H \to \tau \bar \mu)$ as a function of the global Yukawa coupling strength $f$.  The dashed lines are the predictions from the MIA to ${\cal O}(Y_\nu^2+Y_\nu^4)$. The solid lines are the predictions from the full-one loop computation of the mass basis. Here the examples TM4 (left panel) and TM5 (right panel) with $M_R=500,1000$ GeV, as explained in the text,  are chosen.}
\label{totalcomparisonY2Y4_fTM4}
\end{center}
\end{figure}

Thus, it is clear the necessity of considering terms up to ${\cal O}(Y_{\nu}^{4})$ in the MIA. Now we concentrate on the dependence of the branching ratios with $f$. In figure \ref{totalcomparisonY2Y4_fTM4}, we show the partial width and branching ratio as a function of $f$ for the textures TM4 and TM5 with two different values of $M_R$. 
In the perturbativity range of Yukawa couplings (implying approximately $f\lesssim 3.5$) we find a significant increase in the branching ratios up to ${\cal O}(10^{-4})$ for large $f\sim\mathcal O(2)$. 
However, for such large $f$ values the MIA provides an accurate prediction only for large $M_R$ values, say above 1000 GeV, as can be seen in figure \ref{totalcomparisonY2Y4_fTM4}.
Overall, we can conclude that the results for the MIA form factors  to ${\cal O}(Y_\nu^2+Y_\nu^4)$ work reasonably well for heavy  $M_R$ enough, say above 1 TeV and $f$ values not too large, such that $Y_\nu$ is within the perturbativity region, given by $Y_\nu^2/4\pi <1$. We will discuss more on the MIA validity region in the next section.

%%%%%%%%%%%%%%%%%%%%%%%%%%%%%%%%%%%%%
\section{Computation of the one-loop effective vertex for LFVHD}
\label{computationvertex}
In this section we present our results for the form factors involved in our computation of the LFVHD rates in the large $M_R$ limit. Our final purpose here is to derive a simple expression for these form factors at large $M_R\gg v$ that defines in a compact and useful form the one-loop effective vertex for the LFV interaction of our interest here, namely, the interaction of a Higgs boson with two leptons of different flavor $H\ell_k\ell_m$ with $k \neq m$.  The motivation is clear,  with such a simple expression for the involved effective vertex, one may perform a fast estimate of the LFV Higgs decay rates, for many different input parameter values, mainly for 
$Y_\nu$ and $M_R$, without the need of a diagonalization process to reach the physical neutrino basis, and thus avoiding the computation of the full one-loops in this basis which is by far more computer time consuming. In contrast to the full computation, the use of the one-loop effective vertex will provide the explicit dependence on these relevant model parameters, $Y_\nu$ and $M_R$, therefore the interpretation of the numerical results will be easier. The rapid predicted rates with the simple formula that we propose here have, on the other hand, the virtue of  being ready for an easy test with experimental data. 

In order to reach this simple expression for the effective vertex, valid in the large $M_R\gg v $ regime, we perform a systematic expansion  in powers of  $(v/M_R)$ of the one-loop  MIA  amplitude that we have computed in the previous section.  Generically,  the first order in this expansion is ${\cal O} (v^2/M_R^2)$, the next order is ${\cal O} (v^4/M_R^4)$, etc. The logarithmic dependence with $M_R$ is not expanded but left explicit in this calculation. In the final result for the effective vertex we will be interested just in the leading terms of ${\cal O} (v^2/M_R^2)$ which are by far the dominant ones for sufficiently heavy $M_R$.

We start with the formulas found in the previous section and in Appendix \ref{app:FormFactors} for the one-loop LFVHD form factors in the MIA. Assuming the hierarchy $m_{\ell}\ll m_W, m_H\ll M_R$,  we may first ignore the tiny contributions in our analytical results of eqs.~(\ref{FLtot_op2_Y2})-(\ref{FRtot_op2_Y4}) that come from terms in the sum with factors of the lepton masses. 
This leads to the following compact formula to start with for the total one-loop MIA form factors to ${\cal O}(Y_\nu^2+Y_\nu^4)$,  
\begin{align}
F_{L}^{{\rm MIA} }
&= \frac{1}{32 \pi^{2}} \frac{m_{\ell_k}}{m_{W}} \left(Y_{\nu} Y_{\nu}^{\dagger}\right)^{km} \Big(  \tilde{C}_{0}(p_{2},p_{1},m_{W},0,M_{R}) - B_{0}(0,M_{R},m_{W})    \nonumber \\
&-  2m_{W}^{2} \big( (C_{0}+C_{11}-C_{12})(p_{2},p_{1},m_{W},0,M_{R}) + (C_{11}-C_{12})(p_{2},p_{1},m_{W},M_{R},0) \big)     \nonumber \\
&+  4m_{W}^{4} ( D_{12}-D_{13} )(0,p_{2},p_{1},0,M_{R},m_{W},m_{W})    \nonumber \\
&-  2m_{W}^{2}m_{H}^{2}D_{13}(0,p_{2},p_{1},0,M_{R},m_{W},m_{W}) 
 +2m_{W}^{2} \big(C_{0}+C_{11}-C_{12}\big)(p_{2},p_{1},M_{R},m_{W},m_{W})  \nonumber\\
&+  m_{H}^{2} \big(C_{0}+C_{11}-C_{12}\big)(p_{2},p_{1},M_{R},m_{W},m_{W})  \Big)  \nonumber\\
&+\frac{1}{32 \pi^{2}} \frac{m_{\ell_k}}{m_{W}} \left(Y_{\nu} Y_{\nu}^{\dagger} Y_{\nu} Y_{\nu}^{\dagger} \right)^{km} v^{2} \Big(-2(C_{11}-C_{12})(p_{2},p_{1},m_{W},M_{R},M_{R}) 
   \nonumber \\
&+   \tilde{D}_{0}(p_{2},0,p_{1},m_{W},0,M_{R},M_{R}) +\tilde{D}_{0}(p_{2},p_{1},0,m_{W},0,M_{R},M_{R}) -C_{0}(0,0,M_{R},M_{R},m_{W}) \Big)  , \nonumber\\
&
\label{FLtotdom}
\end{align}
where we have ordered the various contributions as follows: the first line is from diagrams (1)+(8)+(10),  the second line
from (2), the third line from (3), the fourth line from (4)+(5), the fifth line from (6) and the last two lines containing the 
${\cal O}(Y_\nu^4)$ contribution are from (1)+(8)+(10). Notice that there are not final contributions from (7)+(9), and the reason is because the two diagrams cancel each other. Similarly, for the right-handed form factor we get:   
\begin{align}
F_{R}^{{\rm MIA}} &= \frac{1}{32 \pi^{2}} \frac{m_{\ell_m}}{m_{W}} \left(Y_{\nu} Y_{\nu}^{\dagger}\right)^{km} \Big(  \tilde{C}_{0}(p_{2},p_{1},m_{W},M_{R},0) - B_{0}(0,M_{R},m_{W})    \nonumber \\
&  -2m_{W}^{2} ( C_{12}(p_{2},p_{1},m_{W},0,M_{R}) +(C_{0}+C_{12})(p_{2},p_{1},m_{W},M_{R},0) )    \nonumber \\
&  +4m_{W}^{4}  D_{13}(0,p_{2},p_{1},0,M_{R},m_{W},m_{W})    \nonumber \\
&  -2m_{W}^{2}m_{H}^{2}(D_{12}-D_{13})(0,p_{2},p_{1},0,M_{R},m_{W},m_{W}) +2m_{W}^{2} \big(C_{0}+C_{12}\big)(p_{2},p_{1},M_{R},m_{W},m_{W}) \nonumber\\
&  +m_{H}^{2} \big(C_{0}+C_{12}\big)(p_{2},p_{1},M_{R},m_{W},m_{W}) \Big)  \nonumber\\
&+\frac{1}{32 \pi^{2}} \frac{m_{\ell_m}}{m_{W}} \left(Y_{\nu} Y_{\nu}^{\dagger} Y_{\nu} Y_{\nu}^{\dagger} \right)^{km} v^{2} \Big(-2C_{12}(p_{2},p_{1},m_{W},M_{R},M_{R})
     \nonumber \\
&   +\tilde{D}_{0}(p_{2},p_{1},0,m_{W},M_{R},M_{R},0) +\tilde{D}_{0}(p_{2},0,p_{1},m_{W},M_{R},M_{R},0)-C_{0}(0,0,M_{R},M_{R},m_{W})
  \Big) \, , \nonumber\\
&
\label{FRtotdom}
\end{align}
where the explanation for the various contributions in each line is as specified above for $F_L$. 
Note also that the right-handed form factor can be obtained from the left-handed one by exchanging $p_2$ and $m_{\ell_k}$ with $p_3$ and $m_{\ell_m}$. 
From the previous compact formula, since we are assuming the hierarchy $m_{\ell_{k}}\gg m_{\ell_{m}}$, it is also clear that the left-handed form factor is the dominant one for the decay mode $H \to \ell_{k} \bar{\ell}_{m}$. Conversely, the right-handed form factor will be the dominant one in the opposite case $m_{\ell_{m}}\gg m_{\ell_{k}}$. For the rest of this section, we will assume $m_{\ell_{k}}\gg m_{\ell_{m}}$ and, consequently, we will focus on the dominant $F_L$.

The next step is to perform the large $M_R$ expansion of the loop integrals appearing in the MIA form factors. The details of how we perform these expansions  and the results for both the loop integrals and the separate contributions to the form factors from all type of diagrams, i=1..10, are collected in  Appendix \ref{app:LoopIntegrals}. 
Finally, by plugging these large $M_R$ expansions into eq.~(\ref{FLtotdom}) we get the one-loop effective vertex, $F_L\simeq V_{H\ell_{k}\ell_{m}}^{\rm eff}$, which parametrizes the one-loop amplitude of  $H \to \ell_k {\bar \ell}_m$, 
\begin{equation}
i {\cal M} \simeq -i g \bar{u}_{\ell_k} V_{H\ell_{k}\ell_{m}}^{\rm eff} P_L  v_{\ell_m} \, , 
\label{amplitudeVeff}
\end{equation}
and the corresponding partial decay width: 
\be
\Gamma (H \to \ell_{k}\bar{\ell}_{m})\simeq \frac{g^{2}}{16\pi}m_{H}\big\vert V_{H\ell_{k}\ell_{m}}^{\rm eff}\big\vert^{2} \, .
\label{Gammaeff}
\ee
We find  the following simple result for the on-shell Higgs boson effective LFV vertex:
\be
V_{H\ell_{k}\ell_{m}}^{\rm eff}=\frac{1}{64 \pi^{2}} \frac{m_{\ell_k}}{m_{W}}  \left[  \frac{m_{H}^{2}}{M_{R}^{2}}
\left( r\Big(\frac{m_{W}^{2}}{m_{H}^{2}}\Big) +\log\left(\frac{m_{W}^{2}}{M_{R}^{2}}\right) \right) \left(Y_{\nu} Y_{\nu}^{\dagger}\right)^{km} - \frac{3v^{2}} {M_{R}^{2}} \left(Y_{\nu} Y_{\nu}^{\dagger} Y_{\nu} Y_{\nu}^{\dagger} \right)^{km} \right]
\label{Veffsimple} \, ,
\ee
where,
\be
r(\lambda)=-\frac{1}{2} -\lambda -8\lambda^{2} +2(1-2\lambda +8\lambda^{2})\sqrt{4\lambda-1}\arctan\left(\frac{1}{\sqrt{4\lambda-1}}\right) +16\lambda^{2}(1-2\lambda)\arctan^2\left(\frac{1}{\sqrt{4\lambda-1}}\right).
\label{rlambda}
\ee 
 Notice that this solution is valid for $m_H<2m_W$. For the physical values of $m_H=125$~GeV and $m_W=80.4$~GeV we get numerically $r(m_W^2/m_H^2)\sim 0.31$.
 The partial width is then simplified correspondingly to:
\be
\Gamma (H \to \ell_{k}\bar{\ell}_{m})^{\rm MIA}= \frac{g^{2}m_{\ell_k}^{2}m_H}{2^{16} \pi^{5}m_{W}^{2}}  \bigg| \frac{m_{H}^{2}}{M_{R}^{2}}
\left( r\Big(\frac{m_{W}^{2}}{m_{H}^{2}}\Big) +\log\left(\frac{m_{W}^{2}}{M_{R}^{2}}\right) \right) \left(Y_{\nu} Y_{\nu}^{\dagger}\right)^{km} - \frac{3v^{2}}{M_{R}^{2}} \left(Y_{\nu} Y_{\nu}^{\dagger} Y_{\nu} Y_{\nu}^{\dagger} \right)^{km} 
 \bigg|^{2} \, .
\label{widthsimple}
\ee
Some comments are in order. First we notice that the dominant behavior with $M_R$ of $V_{H\ell_{k}\ell_{m}}^{\rm eff}$  for large $M_R$ goes as $\log\left(M_{R}^{2}\right)/M_{R}^{2}$ and the next dominant one goes as $1/M_{R}^{2}$ . Second, the terms of ${\cal O}(Y_{\nu}^{2})$ depend on 
$m_H$, whereas the terms of  ${\cal O}(Y_{\nu}^{4})$ do not. Notice also that the two contributions of  ${\cal O}(Y_{\nu}^{2})$ and ${\cal O}(Y_{\nu}^{4})$ get $M_R^2$ in the denominator and not $M_R^4$ as one could naively expect for the ${\cal O}(Y_{\nu}^{4})$ term. Third, we have also checked that we recover the simple phenomenological formula for the LFVHD branching ratio in the case of large Yukawa couplings that was obtained in \cite{Arganda:2014dta} by a naive numerical fit of the dominant contributions at large $M_R$ from diagrams (1)+(8)+(10) in the physical mass basis.  Specifically, if we extract the contributions exclusively from diagrams (1), (8) and (10) in our MIA results in eq.~(\ref{widthsimple}), we get:
\ba
{\rm BR} (H \to \ell_{k}\bar{\ell}_{m})^{\rm MIA}_{(1)+(8)+(10)}&=& \frac{g^{2}m_{\ell_k}^{2}m_H}{2^{16} \pi^{5}m_{W}^{2}\Gamma_H}   \Big\lvert \frac{m_{H}^{2}}{M_{R}^{2}}
  \left(Y_{\nu} Y_{\nu}^{\dagger}\right)^{km} - \frac{3v^{2}}{M_{R}^{2}} \left(Y_{\nu} Y_{\nu}^{\dagger} Y_{\nu} Y_{\nu}^{\dagger} \right)^{km} 
 \Big\rvert^{2} \, \nonumber \\
 &\simeq & 10^{-7} \frac{v^4}{M_R^4}\Big\lvert \left(Y_{\nu} Y_{\nu}^{\dagger}\right)^{km} -  5.7 \left(Y_{\nu} Y_{\nu}^{\dagger} Y_{\nu} Y_{\nu}^{\dagger} \right)^{km} 
\Big\rvert^{2},
\label{BR1810}
\ea
where in the last line we have used the numerical values of the physical parameters with $m_H=125$ GeV and $\Gamma_H$ given by the predicted value in the SM.  As announced,  we reach the same result in eq.~(31) as that given in \cite{Arganda:2014dta} which was derived in the physical neutrino basis. 

It is also illustrative to compare our previous MIA results for the partial width in the large $M_R$ regime with the analytical approximate formula that was found in \cite{Pilaftsis:1992st} from a full one-loop computation in the physical neutrino mass basis and where an expansion in inverse powers of the physical heavy neutrino mass $m_N$ was performed. Concretely  we compare our result in eq.~(\ref{widthsimple}) with those in eqs.~(26), (31)-(34) of \cite{Pilaftsis:1992st}, which were obtained assuming $m_H^2/m_W^2\ll 4$ and $m_H^2/m_N^2\ll1$. After doing some algebra to express the couplings of the physical states appearing in those equations, $B_{\ell_ij}$ and $C_{ij}$,  in terms of the Yukawa couplings, which can be easily done in the seesaw limit,  $m_D\ll m_N$, after neglecting ${\cal O}(1/m_N^4)$ and higher order terms, and finally by extracting just the $m_H$ independent terms, we get from \cite{Pilaftsis:1992st}, a coincident result with our formula in eq.~(\ref{widthsimple}), but when the effective vertex is evaluated in the $m_H \to 0$  limit. Specifically, by using 
\begin{align}
\sum_{i\in{\rm Heavy}} B_{\ell_ki}B^*_{\ell_mi}&\simeq \frac{v^2}{m_N^2} \big(Y_\nu Y_\nu^\dagger\big)^{km} \,,\\
\sum_{i,j\in{\rm Heavy}} B_{\ell_ki}C_{ij}B^*_{\ell_mj}&\simeq \frac{v^4}{m_N^4} \big(Y_\nu Y_\nu^\dagger Y_\nu Y_\nu^\dagger\big)^{km} \,,
\end{align}
we get from \cite{Pilaftsis:1992st}, 
\be
\Gamma (H \to \ell_{k}\bar{\ell}_{m})^{\rm full} \simeq \frac{g^{2}m_{\ell_k}^{2}m_H}{2^{16} \pi^{5}m_{W}^{2}}  \Big\lvert  -\frac{3m_W^2}{m_N^2}  \left(Y_{\nu} Y_{\nu}^{\dagger}\right)^{km} - \frac{3v^{2}}{m_{N}^{2}} \left(Y_{\nu} Y_{\nu}^{\dagger} Y_{\nu} Y_{\nu}^{\dagger} \right)^{km} 
 \Big\rvert^{2} \, ,
\label{widthmhzero}
\ee
which matches with our result in eq.~(\ref{widthsimple}) after the substitution $(m_H^2/m_W^2) r(m_W^2/m_H^2) \to -3$ corresponding to the limit $m_H \to 0$. 
In this sense, although a complete comparison is out of the scope of this work, we conclude that our MIA effective vertex and the effective vertex of the mass basis in \cite{Pilaftsis:1992st} agree analytically in the  limit $m_H \to 0$.  
In any case, we have checked by a numerical estimate of the LFVHD widths that the approximation of neglecting the Higgs boson mass in the effective vertex does not provide an accurate result and, therefore, in order to obtain a realistic estimate of these branching ratios, our final formula for the effective vertex in eq.~(\ref{Veffsimple}) should be used, which is specific for on-shell Higgs decays and accounts properly for the Higgs boson mass effects. 

\begin{figure}[t!]
\begin{center}
\includegraphics[width=.49\textwidth]{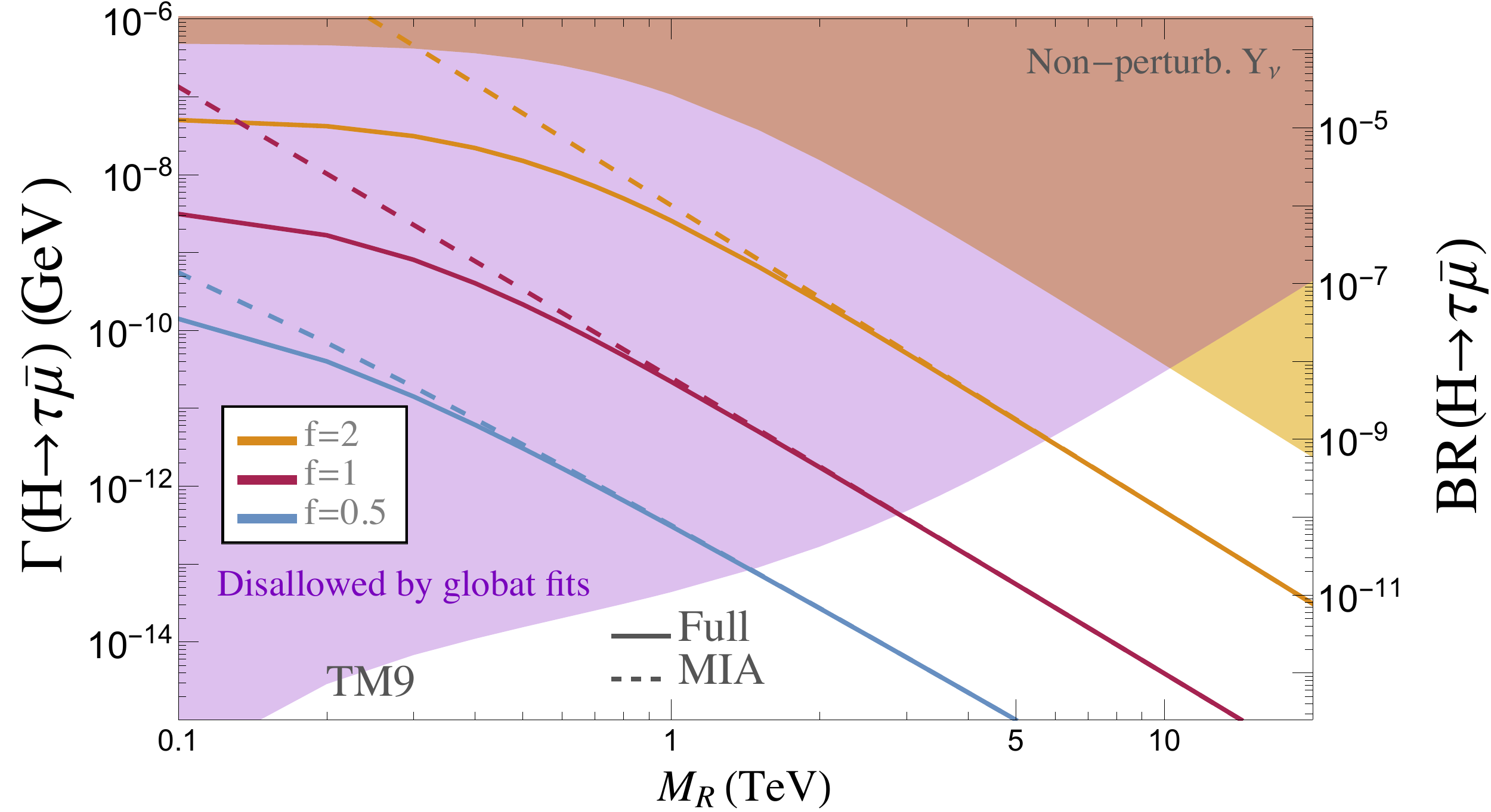}
\includegraphics[width=.49\textwidth]{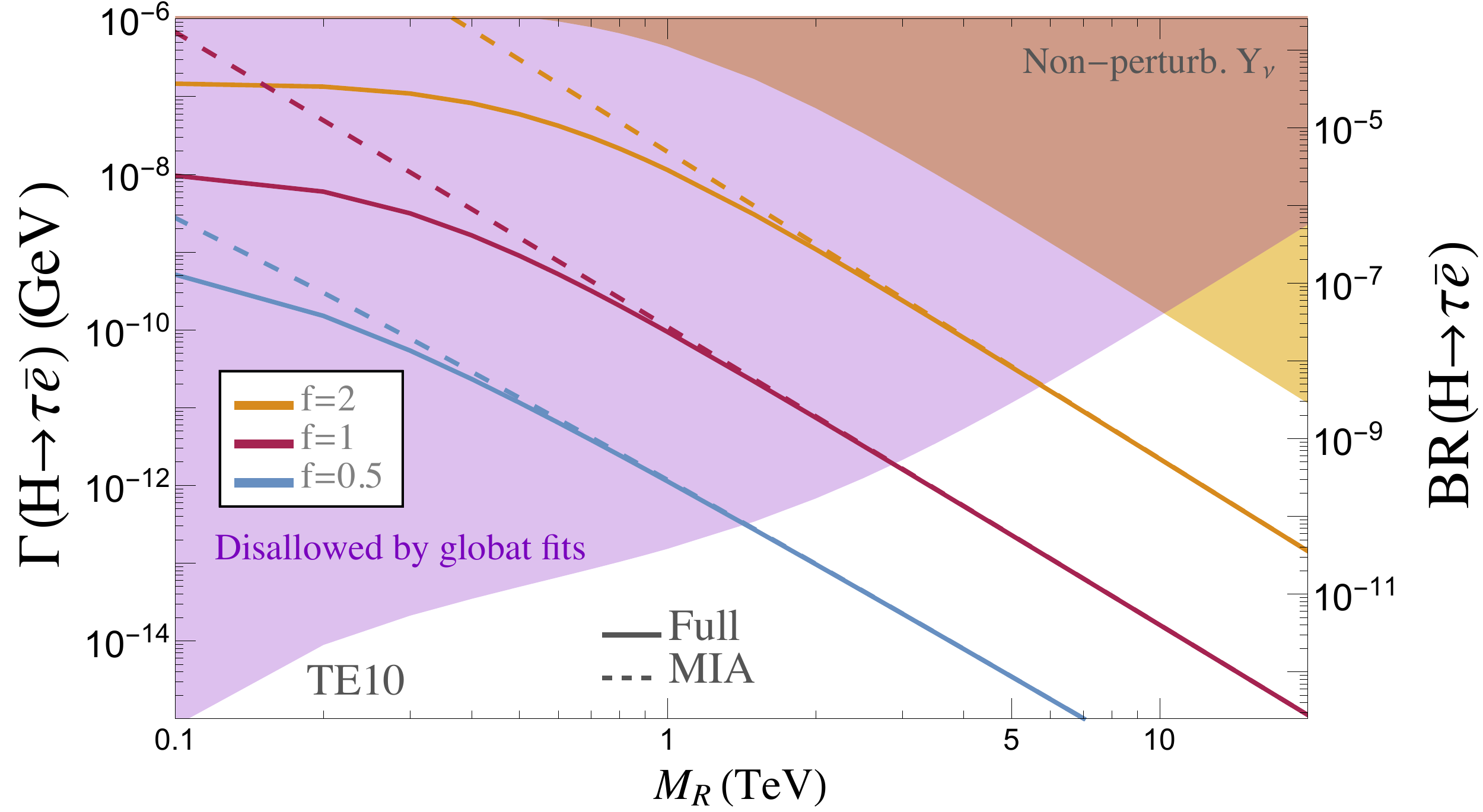}
\caption{Left panel: Predictions for $H \to \tau \bar \mu$ with the effective vertex computed with the MIA (dashed lines) for $Y_\nu^{\rm TM9}$. Right panel:
Predictions for $H \to \tau \bar e$ with the effective vertex computed with the MIA (dashed lines) for $Y_\nu^{\rm TE10}$. The chosen examples TM9 and TE10 are explained in the text. Solid lines are the corresponding predictions from the full one-loop computation in the mass basis. Shadowed areas to the left part of these plots (in purple) are disallowed by global fits. Shadowed areas to the right part of these plots (in yellow) give a nonperturbative Yukawa coupling. }
\label{effvTM9TE10}
\end{center}
\end{figure}

\begin{figure}[t!]
\begin{center}
\includegraphics[width=.49\textwidth]{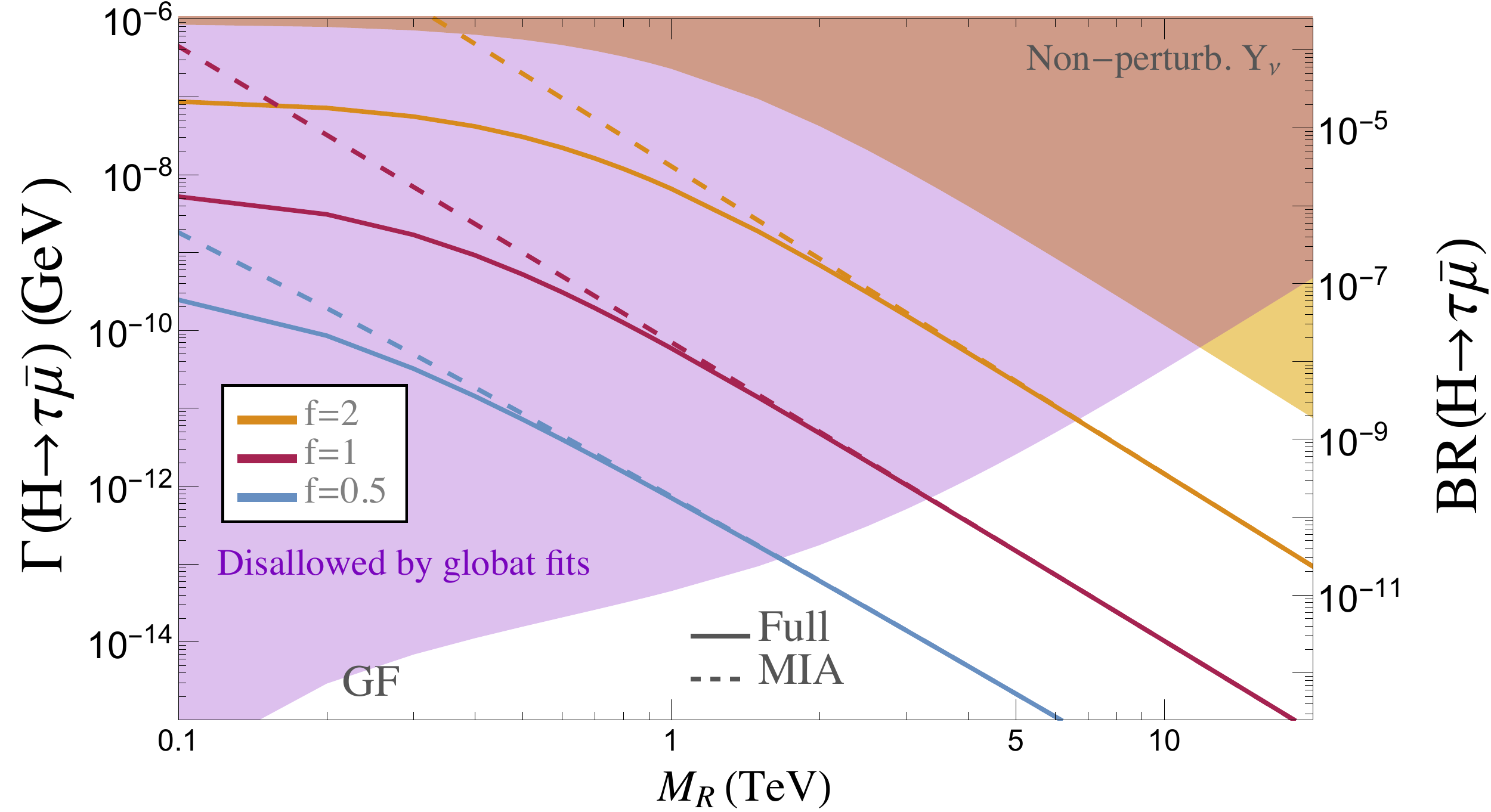}
\includegraphics[width=.49\textwidth]{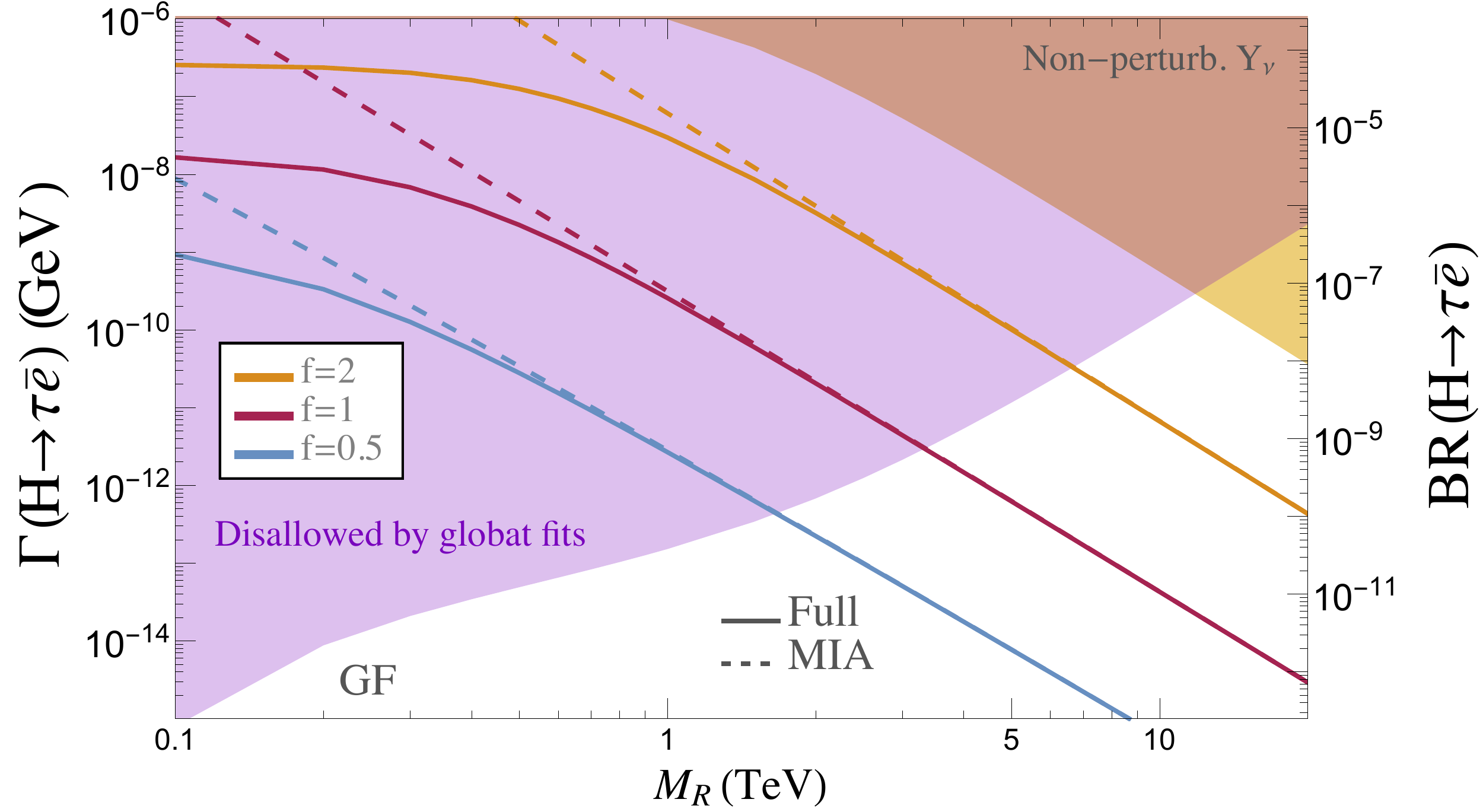}
\caption{Predictions for $H \to \tau \bar \mu$ (left panel) and 
$H \to \tau \bar e$ (right panel) with the effective vertex computed with the MIA (dashed lines) for $Y_\nu^{\rm GF}$. The chosen example GF is explained in the text. Solid lines are the corresponding predictions from the full one-loop computation in the mass basis. Shadowed areas to the left part of these plots (in purple) are disallowed by global fits. Shadowed areas to the right part of these plots (in yellow) give a nonperturbative Yukawa coupling.}
\label{effvGF}
\end{center}
\end{figure}

Finally, we wish to illustrate numerically the goodness of our simple results of the effective vertex and the partial width in eqs.~(\ref{Veffsimple}) and (\ref{widthsimple}), respectively. For that purpose, we compare again our numerical predictions from these simple formulas  with the predictions from the full one-loop results of the mass basis in \cite{Arganda:2014dta}. We do this comparison for the three examples, $Y_\nu^{\rm TM9}$,  $Y_\nu^{\rm TE10}$ and $Y_\nu^{\rm GF}$ in eqs.(\ref{YnuTMTE}) and (\ref{YnuGF}). We show in figures \ref{effvTM9TE10} and \ref{effvGF} this comparison for the most interesting channels $H \to \tau \bar\mu$ and $H \to \tau \bar e$. We have also computed the other channel, $H \to \mu \bar e$, but we do not show here the predictions in this case, because the rates are extremely tiny, therefore irrelevant for phenomenology. The plots in figures \ref{effvTM9TE10} and \ref{effvGF} show the predictions of both the LFVHD partial widths and branching ratios, as functions of $M_R$ and three different values of $f=2,1,0.5$. We have also included the areas that are disallowed by present data as extracted from the analysis with global fits. 
Concretely, we have imposed the constraints on the parameter $\eta$ that we have taken from \cite{Fernandez-Martinez:2016lgt} at the three sigma level. Correspondingly, the shadowed area (in purple) in these plots correspond to values of the parameter space, ($M_R$, $f$) , where the predicted $\eta$ matrix violates the constraint provided by $\eta_{3\sigma}^{\rm max}$ in eq.~(\ref{etamax3sigma}) (at least in one entry).  We have also included in these plots the areas where we do not trust the predictions because  the Yukawa coupling matrix becomes nonperturbative. Specifically, the shadowed area (in yellow) signals $|Y_\nu^{ij}|^2/(4 \pi) >1 $ for some entry. The dark areas (in brown) are the intersections of the previous purple and yellow areas. The areas in white are in consequence the regions that are allowed by the global fits and by perturbativity. 

From these plots the conclusions are immediate and clear. The
agreement between the full prediction and the MIA result obtained from the
effective vertices computed in this section is quite good for values of $M_R$
above 1 TeV and for all the explored Yukawa coupling examples.
 In fact, the MIA works extremely well for the whole region of interest where both the global fits constraints and the perturbativity of the Yukawa coupling are respected. This happens in the white regions of these plots. Last but not least, the predicted LFVHD rates in this allowed region gives, for the three examples analyzed in this section,  maximum branching ratio values of about BR$(H \to \tau \bar\mu) \sim 10^{-8}$ and  BR$(H \to \tau \bar e) \sim 10^{-7}$, far from the present sensitivity which is close to $10^{-2}$.

\section{Conclusions}
\label{conclusions}
In this paper we have computed the LFVHD rates that are induced radiatively to one-loop level from the right-handed neutrinos within the context of the ISS model which is one example of low scale seesaw models with an interesting phenomenology. The full one-loop computation of the partial widths $\Gamma (H \to \ell_i \bar \ell_j)$ in the ISS model was done previously in \cite{Arganda:2014dta} but in the present paper we perform this computation using a very different approach which turns out to provide simpler and more useful analytical results. Instead of applying the usual diagrammatic method of  the full one-loop computation, we have used the MIA  which works with the chiral EW neutrino basis, including the left- and right-handed states $\nu_L$, and $\nu_R$ and the extra singlets, $X$  of the ISS,  instead of dealing with the nine physical neutrino states, $n_i$ $(i=1,..9)$ of the mass basis.   
In order to simplify further this MIA computation we have first prepared the chiral basis in a convenient way, such that all the effects of the singlet $X$ states are collected into a redefinition of the $\nu_R$ propagator,  which we have called here {\it fat propagator},  and then we have derived the set of Feynman rules for these proper chiral states that summarizes the relevant interactions involved in the computation of the LFVHD rates. The peculiarity of using this particular chiral basis is that it leads to a quite generic set of Feynman rules for the subset of interactions involving the neutrino sector, mainly $\nu_L$ and $\nu_R$,  which are the relevant ones for the LFV observables of our interest here, and these could be valid for other low scale seesaw models sharing these same Feynman rules.  With the MIA  we have then organized the one-loop computation of the LFVHD rates in terms of a perturbative expansion in powers of the unique neutrino coupling which may change lepton flavor,  the Yukawa coupling $Y_\nu$. It is worth recalling that the ISS model, under the hypothesis considered here of diagonal $M_R$,  contains two mass insertions leading to lepton flavor change, one is $m_D=vY_\nu$ and the other one is the lepton number violating matrix $\mu_X$. This later is the responsible one for the observed light neutrino masses within the ISS model, and they are generically  flavor nondiagonal. However, due to its tiny size, as to explain the small size of the neutrino masses,  their effects in the LFVHD rates are totally irrelevant and can be safely ignored. Thus the MIA expansion parameter is the $Y_\nu$ matrix being generically non diagonal in flavor,  and it is the unique relevant origin of LFV in our framework.

We have presented here the analytical results using the MIA for the form factors that define the one-loop LFVHD amplitude, and we have done this computation first to  leading order, ${\cal O}((Y_\nu Y_\nu^\dagger)_{ij})$, and later to the next to leading order, i.e. including terms up to ${\cal O}((Y_\nu Y_\nu^\dagger Y_\nu Y_\nu^\dagger)_{ij})$.  Indeed we have demonstrated that our analytical results are gauge invariant. Concretely, we have got the same results of the form factors in eqs. (\ref{FLtotdom}) and (\ref{FRtotdom}) by doing the computation of the form factors in two different gauges: the Feynman-'t Hooft gauge and the unitary gauge.  This is certainly a good check of our analytical results.
 
The most important new analytical results regarding the effective LFV $H\ell_i\ell_j$ vertex are summarized in eqs. (\ref{amplitudeVeff})-(\ref{rlambda}) and those regarding the corresponding partial width are summarized in eq. (\ref{widthsimple}). As it can be seen in these equations, our results contain the generic $Y_\nu$ and do not depend on specific assumptions on this coupling. The only condition it must fulfill for our expansions to be valid is basically that $v Y_\nu \ll M_R$, i.e. that the Dirac mass be smaller than the right-handed neutrino mass scale. Therefore, our analytical results presented here are general within the ISS, are given in terms of just $Y_\nu$ and $M_R$, and they can be applied to other similar low scale seesaw models.
 
We have then explored  the goodness of the MIA results by comparing their corresponding numerical predictions for the partial widths and branching ratios of the $H \to \ell_i \bar \ell_j$ decays with those from the full one-loop computation in the mass basis, which we take from \cite{Arganda:2004bz} and \cite{Arganda:2014dta}. We have found out that in order to get a good numerical convergence of the MIA with the full results, it is absolutely necessary to include both terms, i.e. ${\cal O}(Y^2+Y^4)$ in short,  in the expansion. We have then checked numerically that the MIA works pretty well in a big range of the relevant model parameters, $Y_\nu$ and $M_R$. For a small Yukawa coupling,  given in our notation  by a small global factor say $f<0.5$ we have got an extremely good convergence even for moderate $M_R$ of a few hundred GeV and above. For larger Yukawa couplings, say with $0.5< f< 2$ we have also found a good convergence, but for heavier $M_R$, say above 
${\cal O}(1 {\rm TeV})$. 
 
 In addition to the form factors, we have also derived in this work using the MIA the analytical results of the LFV effective vertex describing the $H\ell_i\ell_j$ coupling that is radiatively generated to one-loop from the  heavy right-handed neutrinos. For that computation we have presented our systematic expansion of the form factors in inverse powers of $M_R$, which is valid in the mass range of our interest, $m_\ell \ll m_D,m_W,m_H\ll M_R$, and we have found the most relevant terms of ${\cal O}(v^2/M_R^2)$ in this series. In doing this expansion, we have taken care of the contributions from the external Higgs boson momentum which are relevant since in this observable the Higgs particle is on-shell,   and we  have also followed the track of all the EW masses involved  like $m_W$ and $m_H$ which are both of order 
 $v$ and therefore contribute to the wanted ${\cal O}(v^2/M_R^2)$ terms.  The lepton masses (except for the global factor) do not provide relevant corrections and have been neglected in this computation of the effective vertex. 
 We believe that our final analytical formula for the LFV effective $H\ell_i\ell_j$ vertex given in eq. (\ref{Veffsimple})  is very simple and can be useful for other authors who wish to perform a fast estimate of the LFVHD rates in terms of their own preferred parameter input values, $Y_\nu$  and $M_R$. We have shown with several examples that this simple MIA formula works extremely well for the interesting window in the $(Y_\nu, M_R)$ parameter space which is allowed by the present experiments. The predicted ratios, BR($H \to \tau \bar \mu$) and BR($H \to \tau \bar e$),  for these particular examples that we have chosen in this work turn out to be significantly constrained by the global fits to present data and by the perturbativity requirements on the Yukawa couplings. These constraints result in allowed LFVHD ratios being at most at about $10^{-7}$ which are unfortunately far below the present experimental sensitivity.

\section*{Acknowledgments}
This work is supported by the European Union through the ITN ELUSIVES H2020-MSCA-ITN-2015//674896 and the RISE INVISIBLESPLUS H2020-MSCA-RISE-2015//690575, by the CICYT through the projects FPA2012-31880 and FPA2016-78645-P (MINECO/FEDER, EU), by the Spanish Consolider-Ingenio 2010 Programme CPAN (CSD2007-00042) and by the Spanish MINECO's ``Centro de Excelencia Severo Ochoa''  Programme under Grant No. SEV-2012-0249. This work was partially supported by ANPCyT PICT 2013-2266 (A.S., E.A., and R.M.). X. M. is supported through the FPU Grant No. AP-2012-6708.

%%%%%%%%%%%%%%%%%%%%%%%%%%%%%%%%%%%%%%
%%%%%%%%%%%%%%%%%%%%%%%%%%%%%%%%%%%%%%

%%%%%%%%%%%%%%%%%%%%%%%%%%%%%%%%%%%%%%
%\newpage
\begin{appendix}
\section{Appendix: Modified neutrino propagators} 
\label{Propagators}

In this appendix we derive the right-handed neutrino  {\it fat propagators} used for the computations in this work. The idea is to resum all possible large flavor diagonal $M_R$ mass insertions, which we denote with a dot in order to distinguish them from the flavor off-diagonal ones, in a way such that the large mass appears effectively in the denominator of the propagators of the new states. 

\begin{figure}[b!]
\begin{center}
\includegraphics[width=\textwidth]{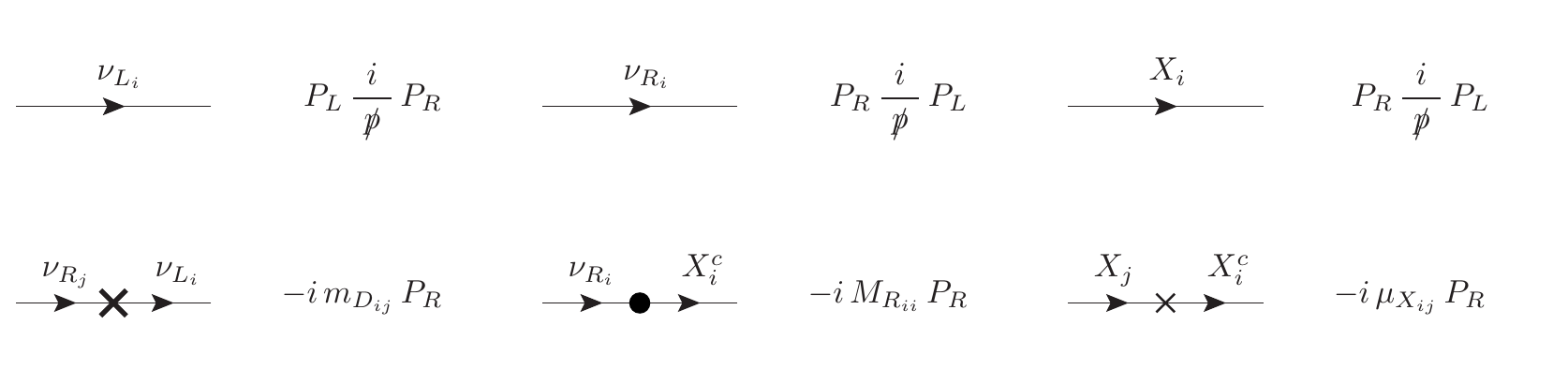}
\caption{Propagators and mass insertions in the electroweak basis. Crosses denote mass insertions that can change flavor, while big black dots are for flavor diagonal ones.}\label{PropsandInsertions}
\end{center}
\end{figure}

In order to make a MIA computation in the electroweak basis $(\nu_L^c\,,\;\nu_R\,,\;X)$, we need to take into account all the propagators and mass insertions given by the neutrino mass matrix. In the ISS model we are considering, this mass matrix is given by eq.~(\ref{ISSmatrix}), which we repeat here for completeness:
\begin{equation}
\label{ISSmatrixApp}
 M_{\mathrm{ISS}}=\left(\begin{array}{c c c} 0 & m_D & 0 \\ m_D^T & 0 & M_R \\ 0 & M_R^T & \mu_X \end{array}\right)\,.
\end{equation}
From this mass matrix, we obtain the propagators and mass insertions summarized in figure~\ref{PropsandInsertions}.
It is important to notice the presence of the $P_L$ and $P_R$ projectors for the chiral fields, which have been properly added according to: 
\begin{equation}
\begin{array}{lcl}
\nu_L^c , \nu_R , X  &\longrightarrow& {\rm \bf RH~ fields} , \\
\nu_L , \nu_R^c , X^c  &\longrightarrow& {\rm \bf LH~ fields} .
\end{array}
\end{equation}
As previously mentioned, there are three types of mass insertions and they are controlled by the matrices $m_D$, $M_R$ and $\mu_X$. 
The mass insertions $M_R$ that relate $\nu_R$ and $X$ fields are taken to be flavor diagonal in this work and are denoted by a big dot in figure~\ref{PropsandInsertions}.
On the other hand, crosses indicate flavor nondiagonal insertions coming from $m_D$ (big cross) and $\mu_X$ (small cross), which connect the fields $\nu_L$-$\nu_R$ and two $X$'s respectively.
Nevertheless, given that we work under the assumption that $\mu_X$ is a tiny scale, we neglect $\mu_X$ mass insertions for our LFVHD computations and, therefore, we consider $m_D$ as the only relevant LFV insertion.

\begin{figure}[t!]
\begin{center}
\includegraphics[width=\textwidth]{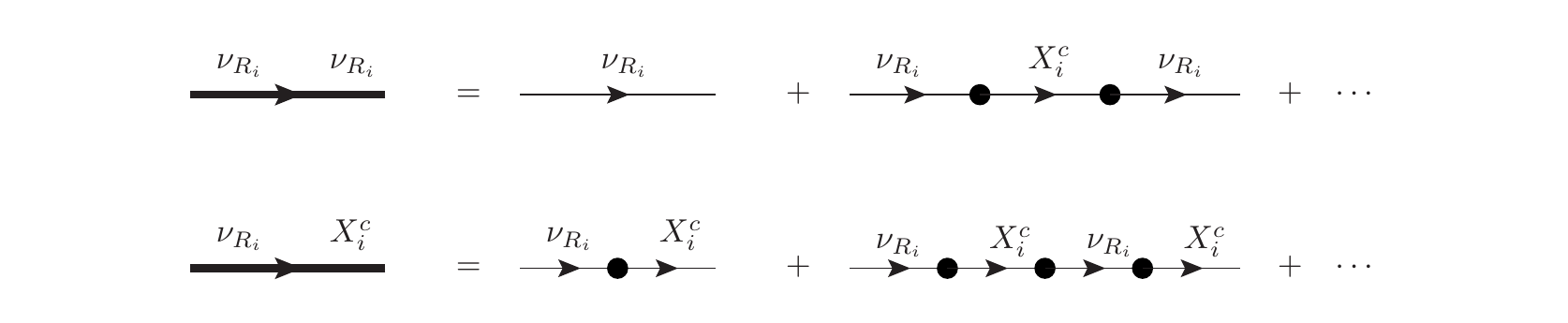}
\caption{Modified neutrino propagators after resuming an infinite number of $M_R$ mass insertions, denoted here by big black dots. We use fat arrow lines  with same (different) initial and final states to denote that all possible even (odd) number of $M_R$ insertions have been considered.  The fat lines with same initial and final $\nu_R$ states are referred to in this work as {\it fat propagators}.}\label{FatProps}
\end{center}
\end{figure}

Since our motivation in this work is to make a MIA computation for LFV H decays by perturbatively inserting LFV mass insertions, we find convenient to take into account first the effects of all possible flavor diagonal $M_R$ insertions. 
Moreover, this procedure allows us to consider $M_R$ also as a heavy scale so we can define an effective vertex for the $H$-$\ell_i$-$\ell_j$ interaction.
This can be done by defining two types of modified propagators, one for same initial and final state consisting of all possible even number of $M_R$ insertions (which we call {\it fat propagator}), and one for different initial and final states with an odd number of $M_R$ insertions, as it is schematically shown in figure~\ref{FatProps}.
We can then define two modified propagators starting with $\nu_R$ by adding the corresponding series:
\begin{align}
{\rm Prop}_{\, \nu_{R_i}\to\nu_{R_i}}
&= P_R \dfrac i{\slashed p}  P_L + P_R \dfrac i{\slashed p}  P_L ~ \Big(-i M_{R_i}^* P_L \Big)~ P_L \dfrac i{\slashed p}  P_R ~ \Big(-i M_{R_i} P_R \Big)~ P_R \dfrac i{\slashed p}  P_L + \cdots\nonumber\\
& = P_R~ \dfrac i{\slashed p}\sum_{n\geq0}\bigg(\dfrac{|M_{R_i}|^2}{p^2}\bigg)^n ~ P_L= P_R~ \dfrac{ i\slashed p}{p^2-|M_{R_i}|^2}~ P_L\,, \label{FatpropnnuR}
\\\nonumber\\
{\rm Prop}_{\, \nu_{R_i}\to X_{i}^c}
&= P_L \dfrac i{\slashed p}  P_R ~  \Big(-i M_{R_i} P_R \Big)~ P_R \dfrac i{\slashed p}  P_L
\nonumber\\
&+ ~ P_L \dfrac i{\slashed p}  P_R ~  \Big(-i M_{R_i} P_R \Big)~ P_R \dfrac i{\slashed p}  P_L~   \Big(-i M_{R_i}^* P_L \Big)~ P_L \dfrac i{\slashed p}  P_R ~  \Big(-i M_{R_i} P_R \Big)~ P_R \dfrac i{\slashed p}  P_L + \cdots \nonumber\\
&= P_L~ \dfrac {i M_{R_i}}{p^2}\sum_{n\geq0}\bigg(\dfrac{|M_{R_i}|^2}{p^2}\bigg)^n ~ P_L= P_L~ \dfrac{ i M_{R_i}}{p^2-|M_{R_i}|^2}~ P_L\,.
\end{align}
And we can similarly define other modified propagators considering also the $\nu_R^c$ and $X$ states. In the present study of LFVHD, it happens that the $X$ fields do not interact with any of the external legs involved in the LFV process we want to compute. Consequently, to take into account the effects from $X$ in the LFVHD, it is enough to consider the {\it fat propagator} in eq.~(\ref{FatpropnnuR}) when computing the one-loop contributions to $H\to\ell_k\bar\ell_m$.

%%%%%%%%%%%

%%%%%%%%%%%%%%%%%%%%%%%%%%%%%%%%%%%%%%
\section{Appendix: MIA Form Factors (Feynman-'t Hooft gauge)} 
\label{app:FormFactors}

Here we present the analytical results for the form factors $F_{L,R}$ involved in the computation of the LFVHD decay rates when computed with the MIA to one-loop order and considering the leading order corrections, ${\cal O} (Y_\nu^2)$, and the next to leading corrections, ${\cal O} (Y_\nu^4)$, as explained in the text. This means the computation of all the one-loop diagrams in figures \ref{diagsvertexY2}, \ref{diagslegY2}, \ref{diagsvertexY4} and \ref{diagslegY4}. They are written in terms of the usual one-loop functions for the two-point $B's$,  three-point $C's$, and four-point $D's$ functions. 
We follow these definitions and conventions:
\begin{equation}\label{loopfunctionB}
    \mu^{4-D}~ \int \frac{d^D k}{(2\pi)^D} \frac{\{1; k^{\mu}\}}
    {[k^2 - m_1^2][(k+p_1)^2 - m_2^2]}
    = \frac{i}{16\pi^2} \left\{ B_0; p_1^{\mu} B_1 \right\}
    (p_1,m_1,m_2)\,,
\end{equation}
\begin{align}
    \mu^{4-D}& \int \frac{d^D k}{(2\pi)^D}
    \frac{\{1; k^2; k^{\mu}\}}
    {[k^2 - m_1^2][(k + p_1)^2 - m_2^2][(k + p_1 + p_2)^2 - m_3^2]}
    \nonumber \\
     & = \frac{i}{16\pi^2}
    \left\{ C_0; {\tilde C}_0; p_1^{\mu}C_{11} + p_2^{\mu}C_{12} \right\}
    (p_1, p_2, m_1, m_2, m_3)\,, \label{loopfunctionC}
\end{align}
\begin{align}
   \mu^{4-D}& \int \frac{d^D k}{(2\pi)^D}
    \frac{\{1; k^2; k^{\mu}\}}
    {[k^2 - m_1^2][(k + p_1)^2 - m_2^2][(k + p_1 + p_2)^2 - m_3^2][(k + p_1 + p_2+ p_3)^2 - m_4^2]}
    \nonumber \\
     & = \frac{i}{16\pi^2}
    \left\{ D_0; {\tilde D}_0; p_1^{\mu}D_{11} + p_2^{\mu}D_{12} +p_3^{\mu}D_{13}\right\}
    (p_1, p_2, p_3, m_1, m_2, m_3,m_4)\,. \label{loopfunctionD}
\end{align}

We start with the left-handed form factors and present the contributions diagram by diagram following the notation explained in the text and $m_{k,m}\equiv m_{\ell_{k,m}}$.
We will restrict ourselves to the dominant contributions, meaning those that will provide $\mathcal O(v^2/M_R^2)$ terms when doing the large $M_R$ expansion, as explained in the next appendix. 
For instance, contributions from loop functions of type $D_i$ where $M_R$ appears in two of the mass arguments go as $1/M_R^4$ and provide subdominant corrections that are not considered here. 

The results of the ${\cal O} (Y_\nu^2)$ contributions are
%%%%%%%%%%%%%%%
\begin{align}
 F_{L}^{{\rm MIA (1)\,\, (Y^2)}}=& \frac{1}{32 \pi^{2}} \frac{m_{k}}{m_{W}}   \left(Y_{\nu} Y_{\nu}^{\dagger}  \right)^{km} \Big( (\tilde{C}_{0}+m_{k}^{2}(C_{11}-C_{12})+m_{m}^{2}C_{12})_{(1a)}   \nonumber \\
& +m_{m}^{2}(C_{0}+C_{11})_{(1b)} -m_{m}^{2}(C_{12})_{(1c)} -m_{m}^{2}(C_{0}+C_{12})_{(1d)} \Big)     \,,\nonumber \\
%%%
F_{L}^{{\rm MIA (2)\,\, (Y^2)}}=& \frac{-1}{16 \pi^{2}} m_{k}m_{W} \left(Y_{\nu} Y_{\nu}^{\dagger} \right)^{km} \Big( (C_{0}+C_{11}-C_{12})_{(2a)} +(C_{11}-C_{12})_{(2b)} \Big) \,,\nonumber \\
%%%
 F_{L}^{{\rm MIA (3)\,\, (Y^2)}}=& \frac{1}{8 \pi^{2}} m_{k}m_{W}^{3} \left(Y_{\nu} Y_{\nu}^{\dagger} \right)^{km} ( D_{12}-D_{13} )_{(3a)} \,,\nonumber \\
%%%
 F_{L}^{{\rm MIA (4)\,\, (Y^2)}}=& \frac{-1}{32 \pi^{2}} m_{k}m_{W} \left(Y_{\nu} Y_{\nu}^{\dagger} \right)^{km} \Big( (C_{0}-C_{11}+C_{12})_{(4a)} +m_{m}^{2}(2D_{12}-D_{13})_{(4b)} \Big)\,, \nonumber \\
%%%
 F_{L}^{{\rm MIA (5)\,\, (Y^2)}}=& \frac{1}{32 \pi^{2}} m_{k}m_{W}  \left(Y_{\nu} Y_{\nu}^{\dagger}  \right)^{km} \Big( (2C_{0}+C_{11}-C_{12})_{(5a)} \nonumber\\
 &+(C_{0}+2m_{k}^{2}D_{12}-(2m_{H}^{2}-m_{m}^{2})D_{13})_{(5b)} \Big) \,,\nonumber \\
%%%
 F_{L}^{{\rm MIA (6)\,\, (Y^2)}}=& \frac{1}{32 \pi^{2}} \frac{m_{k}}{m_{W}} m_{H}^{2} \left( Y_{\nu} Y_{\nu}^{\dagger} \right)^{km} \Big( (C_{11}-C_{12})_{(6a)} +(C_{0})_{(6c)} +m_{m}^{2}(D_{13})_{(6d)} \Big) \,,\nonumber \\
%%%
 F_{L}^{{\rm MIA (7)\,\, (Y^2)}}=& \frac{1}{16 \pi^{2}} m_{k}m_{W} \frac{m_{m}^{2}}{m_{k}^{2}-m_{m}^{2}}  \left(Y_{\nu} Y_{\nu}^{\dagger} \right)^{km} ( C_{12} )_{(7a)} \,, \nonumber \\
%%%
  F_{L}^{{\rm MIA (8)\,\, (Y^2)}}=& \frac{1}{32 \pi^{2}} \frac{m_{k}}{m_{W}} \frac{m_{m}^{2}}{m_{k}^{2}-m_{m}^{2}}  \left( Y_{\nu} Y_{\nu}^{\dagger} \right)^{km} \Big( (B_{1})_{(8a)} +(B_{0})_{(8b)} +(B_{0})_{(8c)} +m_{k}^{2}(C_{12})_{(8d)} \Big)\,, \nonumber \\
%%%
 F_{L}^{{\rm MIA (9)\,\, (Y^2)}}=& \frac{-1}{16 \pi^{2}} m_{k}m_{W} \frac{m_{m}^{2}}{m_{k}^{2}-m_{m}^{2}}  \left(Y_{\nu} Y_{\nu}^{\dagger} \right)^{km} ( C_{12} )_{(9a)}\,, \nonumber \\
%%%
 F_{L}^{{\rm MIA (10)\,\, (Y^2)}}=& \frac{-1}{32 \pi^{2}} \frac{m_{k}}{m_{W}} \frac{m_{m}^{2}}{m_{k}^{2}-m_{m}^{2}}  \left( Y_{\nu} Y_{\nu}^{\dagger} \right)^{km} \Big( (B_{1})_{(10a)} +(B_{0})_{(10b)} +\frac{m_{k}^{2}}{m_{m}^{2}}(B_{0})_{(10c)} +m_{k}^{2}(C_{12})_{(10d)} \Big) \,   .
  \nonumber \\
 &
\label{FLtot_op2_Y2}
\end{align}
The results of the dominant ${\cal O} (Y_\nu^4)$ contributions are:
\begin{align}
F_{L}^{{\rm MIA (1)\,\, (Y^4)}}=&  \frac{1}{32 \pi^{2}} \frac{m_{k}}{m_{W}}
\left(Y_{\nu} Y_{\nu}^{\dagger} Y_{\nu} Y_{\nu}^{\dagger} \right)^{km} v^{2} \Big( -(C_{11}-C_{12})_{(1e)} -(C_{11}-C_{12}+C_{0})_{(1f)}    \nonumber \\
&+(\tilde{D}_{0})_{(1g)} +(\tilde{D}_{0})_{(1h)} +(C_{0})_{(1j)} \Big)\,,
\nonumber \\
%%%
 F_{L}^{{\rm MIA (8)\,\, (Y^4)}}=& \frac{1}{32 \pi^{2}} \frac{m_{k}}{m_{W}} \frac{m_{m}^{2}}{m_{k}^{2}-m_{m}^{2}}
 \left( Y_{\nu} Y_{\nu}^{\dagger} Y_{\nu} Y_{\nu}^{\dagger} \right)^{km} v^{2}\Big( (C_{12})_{(8e)}+(C_{0})_{(8f)}+(C_{0})_{(8g)} \Big) \,,  \nonumber \\
%%%
F_{L}^{{\rm MIA (10)\,\, (Y^4)}}=& \frac{-1}{32 \pi^{2}} \frac{m_{k}}{m_{W}} \frac{m_{m}^{2}}{m_{k}^{2}-m_{m}^{2}}
\left( Y_{\nu} Y_{\nu}^{\dagger} Y_{\nu} Y_{\nu}^{\dagger} \right)^{km} v^{2}\Big( (C_{12})_{(10e)}+(C_{0})_{(10f)}+\frac{m_{k}^{2}}{m_{m}^{2}}(C_{0})_{(10g)} \Big) \, .  
\label{FLtot_op2_Y4}
\end{align}
Next, we present the right-handed form factors. The results of the  ${\cal O} (Y_\nu^2)$ contributions  are: 
\begin{align}
 F_{R}^{{\rm MIA (1)\,\, (Y^2)}}=& \frac{1}{32 \pi^{2}} \frac{m_{m}}{m_{W}}   \left(Y_{\nu} Y_{\nu}^{\dagger}  \right)^{km} \Big( m_{k}^{2}(C_{0}+C_{11})_{(1a)} +(\tilde{C}_{0}+m_{k}^{2}(C_{11}-C_{12})+m_{m}^{2}C_{12})_{(1b)}  \nonumber \\
& -m_{k}^{2}(C_{0}+C_{11}-C_{12})_{(1c)} -m_{k}^{2}(C_{11}-C_{12})_{(1d)} \Big)  \,,   \nonumber \\
%%%
F_{R}^{{\rm MIA (2)\,\, (Y^2)}}=& \frac{-1}{16 \pi^{2}} m_{m}m_{W} \left(Y_{\nu} Y_{\nu}^{\dagger} \right)^{km} \Big( (C_{12})_{(2a)} +(C_{0}+C_{12})_{(2b)} \Big)\,, \nonumber \\
%%%
 F_{R}^{{\rm MIA (3)\,\, (Y^2)}}=& \frac{1}{8 \pi^{2}} m_{m}m_{W}^{3} \left(Y_{\nu} Y_{\nu}^{\dagger} \right)^{km} (D_{13})_{(3a)} \,,\nonumber \\
%%% 
F_{R}^{{\rm MIA (4)\,\, (Y^2)}}=& \frac{1}{32 \pi^{2}} m_{m}m_{W}  \left(Y_{\nu} Y_{\nu}^{\dagger}  \right)^{km} \Big( (2C_{0}+C_{12})_{(4a)} \nonumber\\
&+\big(C_{0}+2m_{m}^{2}D_{12}-(2m_{H}^{2}-m_{k}^{2})(D_{12}-D_{13})\big)_{(4b)} \Big) \,,\nonumber \\
%%%
 F_{R}^{{\rm MIA (5)\,\, (Y^2)}}=& \frac{-1}{32 \pi^{2}} m_{m}m_{W} \left(Y_{\nu} Y_{\nu}^{\dagger} \right)^{km} \Big( (C_{0}-C_{12})_{(5a)} +m_{k}^{2}(D_{12}+D_{13})_{(5b)} \Big)\,, \nonumber \\
%%%
 F_{R}^{{\rm MIA (6)\,\, (Y^2)}}=& \frac{1}{32 \pi^{2}} \frac{m_{m}}{m_{W}} m_{H}^{2} \left( Y_{\nu} Y_{\nu}^{\dagger} \right)^{km} \Big( (C_{12})_{(6a)} +(C_{0})_{(6b)} +m_{k}^{2}(D_{12}-D_{13})_{(6d)} \Big) \,,\nonumber \\
%%%
 F_{R}^{{\rm MIA (7)\,\, (Y^2)}}=& \frac{1}{16 \pi^{2}} m_{m}m_{W} \frac{m_{k}^{2}}{m_{k}^{2}-m_{m}^{2}}  \left(Y_{\nu} Y_{\nu}^{\dagger} \right)^{km} ( C_{12} )_{(7a)}  \,,\nonumber \\
%%%
  F_{R}^{{\rm MIA (8)\,\, (Y^2)}}=& \frac{1}{32 \pi^{2}} \frac{m_{m}}{m_{W}} \frac{m_{k}^{2}}{m_{k}^{2}-m_{m}^{2}}  \left( Y_{\nu} Y_{\nu}^{\dagger} \right)^{km} \Big( (B_{1})_{(8a)} +\frac{m_{m}^{2}}{m_{k}^{2}}(B_{0})_{(8b)} +(B_{0})_{(8c)} +m_{m}^{2}(C_{12})_{(8d)} \Big) \,,\nonumber \\
%%%
 F_{R}^{{\rm MIA (9)\,\, (Y^2)}}=& \frac{-1}{16 \pi^{2}} m_{m}m_{W} \frac{m_{k}^{2}}{m_{k}^{2}-m_{m}^{2}}  \left(Y_{\nu} Y_{\nu}^{\dagger} \right)^{km} ( C_{12} )_{(9a)}\,, \nonumber \\
%%%
 F_{R}^{{\rm MIA (10)\,\, (Y^2)}}=& \frac{-1}{32 \pi^{2}} \frac{m_{m}}{m_{W}} \frac{m_{k}^{2}}{m_{k}^{2}-m_{m}^{2}}  \left( Y_{\nu} Y_{\nu}^{\dagger} \right)^{km} \Big( (B_{1})_{(10a)} +(B_{0})_{(10b)} +(B_{0})_{(10c)} +m_{m}^{2}(C_{12})_{(10d)} \Big) \, .   \nonumber \\ 
&\label{FRtot_op2_Y2}
\end{align}
The results of the dominant ${\cal O} (Y_\nu^4)$ contributions are:
\begin{align}
F_{R}^{{\rm MIA (1)\,\, (Y^4)}}=&  \frac{1}{32 \pi^{2}} \frac{m_{m}}{m_{W}}
\left(Y_{\nu} Y_{\nu}^{\dagger} Y_{\nu} Y_{\nu}^{\dagger} \right)^{km} v^{2} \Big( -(C_{0}+C_{12})_{(1e)} -(C_{12})_{(1f)}    \nonumber \\
&+(C_{0})_{(1i)} +(\tilde{D}_{0})_{(1k)} +(\tilde{D}_{0})_{(1l)}  \Big)\,,
\nonumber \\
%%%
 F_{R}^{{\rm MIA (8)\,\, (Y^4)}}=& \frac{1}{32 \pi^{2}} \frac{m_{m}}{m_{W}} \frac{m_{k}^{2}}{m_{k}^{2}-m_{m}^{2}}
 \left( Y_{\nu} Y_{\nu}^{\dagger} Y_{\nu} Y_{\nu}^{\dagger} \right)^{km} v^{2}\Big( (C_{12})_{(8e)}+\frac{m_{m}^{2}}{m_{k}^{2}}(C_{0})_{(8f)}+(C_{0})_{(8g)} \Big) \,,  \nonumber \\
%%%
F_{R}^{{\rm MIA (10)\,\, (Y^4)}}=& \frac{-1}{32 \pi^{2}} \frac{m_{m}}{m_{W}} \frac{m_{k}^{2}}{m_{k}^{2}-m_{m}^{2}}
\left( Y_{\nu} Y_{\nu}^{\dagger} Y_{\nu} Y_{\nu}^{\dagger} \right)^{km} v^{2}\Big( (C_{12})_{(10e)}+(C_{0})_{(10f)}+(C_{0})_{(10g)} \Big) \, .
\label{FRtot_op2_Y4}
\end{align}
The arguments of the above one-loop integrals are the following: 
\be\begin{array}{rll}
 \tilde{C}_{0}, C_{i} =& \tilde{C}_{0}, C_{i} (p_{2},p_{1},m_{W},0,M_{R}) & \text{in } (1a), (1c), (2a) \nonumber \\
 \tilde{C}_{0}, C_{i} =& \tilde{C}_{0}, C_{i} (p_{2},p_{1},m_{W},M_{R},0) & \text{in } (1b), (1d), (2b)  \nonumber \\
 C_{i} =& C_{i} (p_{2},p_{1},m_{W},M_{R},M_{R}) & \text{in } (1e), (1f),(1i), (1j)  \nonumber \\
 \tilde{D}_{0} =& \tilde{D}_{0} (p_{2},0,p_{1},m_{W},0,M_{R},M_{R}) & \text{in } (1g)  \nonumber \\
 \tilde{D}_{0} =& \tilde{D}_{0} (p_{2},p_{1},0,m_{W},0,M_{R},M_{R}) & \text{in } (1h)  \nonumber \\
 \tilde{D}_{0} =& \tilde{D}_{0} (p_{2},p_{1},0,m_{W},M_{R},M_{R},0) & \text{in } (1k)  \nonumber \\
 \tilde{D}_{0} =& \tilde{D}_{0} (p_{2},0,p_{1},m_{W},M_{R},M_{R},0) & \text{in } (1l)  \nonumber \\
 D_{i} =& D_{i} (0,p_{2},p_{1},0,M_{R},m_{W},m_{W}) & \text{in } (3a), (4b), (5b), (6d)  \nonumber\\
 C_{i} =&  C_{i} (p_{2},p_{1},M_{R},m_{W},m_{W}) & \text{in } (4a), (4b), (5a), (5b), (6a), (6b), (6c)  \nonumber \\
 C_{12} =& C_{12} (0,p_{2},0,M_{R},m_{W}) & \text{in } (7a), (8d)  \nonumber \\
 B_{i} =& B_{i} (p_{2},M_{R},m_{W}) & \text{in } (8a), (8b), (8c)  \nonumber \\
 C_{i} =& C_{i} (0,p_{2},M_{R},M_{R},m_{W}) & \text{in } (8e), (8f),(8g)   \nonumber \\
 C_{12} =& C_{12} (0,p_{3},0,M_{R},m_{W}) & \text{in } (9a), (10d)  \nonumber \\
 B_{i} =& B_{i} (p_{3},M_{R},m_{W}) & \text{in } (10a), (10b), (10c)  \nonumber \\
 C_{i} =& C_{i} (0,p_{3},M_{R},M_{R},m_{W}) & \text{in } (10e), (10f),(10g) \,.
\label{argfloops_op2}
\end{array}\ee
We want to remark that the above formulas are valid for the degenerate $M_{R_i}=M_R$ case. Nevertheless, they can be easily generalized to the nondegenerate case by properly including the summation indices. For example, it would be enough to change 
\begin{align}
(Y_\nu Y_\nu^\dagger)^{km} C_\alpha(p_2,p_1,m_W,0,M_R) 	&\rightarrow (Y_\nu^{ka} Y_\nu^{\dagger am}) C_\alpha(p_2,p_1,m_W,0,M_{R_a})\,, \nonumber\\
(Y_\nu Y_\nu^\dagger Y_\nu Y_\nu^\dagger)^{km} C_\alpha(p_2,p_1,m_W,M_R,M_R) &\rightarrow	 (Y_\nu^{ka} Y_\nu^{\dagger ai} Y_\nu^{ib} Y_\nu^{\dagger bm}) C_\alpha(p_2,p_1,m_W,M_{R_a},M_{R_b})\,,
\end{align}
and similarly for all the terms. 

%%%%%%%%%%%%%%%%%%%%%%%%%%%%%%%%%%%%%%%%%%%%%%%%%%%%%%%%%%% %%%%%%%%%%%%%%%%%%%%%%%%%%%%%%%%
%%%%%%%%%%%%%%%%%%%%%%%%%%%%%%%%

\section{Appendix: The large $\boldsymbol{M_R}$ expansion} 
\label{app:LoopIntegrals}

Here we present our analytical results for the loop-functions and form factors involved in our computation of LFVHD rates in the large $M_R$ limit. To reach this limit we perform a systematic expansion of the amplitude in powers of $(v^2/M_R^2)$.  Generically,  the first order in this expansion is ${\cal O} (v^2/M_R^2)$ the next order is ${\cal O} (v^4/M_R^4)$, etc. The logarithmic dependence with $M_R$ is left unexpanded. In the final expansion we will keep just the dominant terms in the form factors of ${\cal O} (v^2/M_R^2)$ which will be shown to be sufficient to describe successfully the final amplitude for LFVHD in the heavy right-handed neutrino mass region of our interest, $M_R \gg v$.

We first calculate the large $M_{R}$ expansions of all one-loop functions and second we plug these expansions in the form factors formulas. To do this, we perform first the integration of Feynman's parameters and next we expand  them for large $M_{R}\gg v$.  Since we have the mass of $W$ boson in the mass argument of the one-loop functions, we cannot take the most used approximation of neglecting external momentum particles (because the mass of the Higgs boson enters here). In fact our expansions presented in this work will apply to the present case of on-shell Higgs boson, i.e. with $p_1^2=m_H^2$ and $m_H$ being the realistic Higgs boson mass. Furthermore, it should be noticed that in principle
there are three very different scales of masses involved in the computation: the lepton sector masses ($m_{\ell_m}$ and $m_{\ell_k}$), the electroweak sector masses ($m_W$ and $m_H$) and the new physics scale $M_R$.
As we said, in a good approximation we can neglect the lepton masses in the one-loop functions at the beginning. However, both electroweak masses $m_W$ and $m_H$ must be retained in order to calculate the ${\cal O} (M_{R}^{-2})$ terms of the one-loop functions. Actually, in practice we consider the vacuum expectation value $v$, which is the common scale entering in both electroweak masses within the SM, and as we said above, we perform a well-defined expansion in powers of an unique dimensionless parameter that is given by the ratio $v^{2}/M_{R}^{2}$.

At the numerical level,  we have checked that all expansions presented in the following are in very good accordance with the numerical results from LoopTools~\cite{Hahn:1998yk}. 
The analytical expansions that we get for the dominant terms of the loop functions, i.e. up to $\mathcal O(M_R^{-2})$, are summarized next:
%%%%%%%
\ba
B_{0} \left(p,M_R,m_W\right) &=& \Delta +1 -\log \Big(\frac{M_{R}^{2}}{\mu^2}\Big) +\frac{m_{W}^{2} \log \left(\frac{m_{W}^{2}}{M_{R}^{2}}\right)}{M_{R}^{2}} +\frac{p^2}{2M_R^2}  \,,\nonumber\\
C_{0} \left(p_2,p_1,m_W,0,M_R\right) &=& C_{0} \left(p_2,p_1,m_W,M_R,0\right) = \frac{\log \left(\frac{m_{W}^{2}}{M_{R}^{2}}\right)}{M_{R}^{2}}  \,, \nonumber\\
C_{0} \left(p_2,p_1,M_{R,}m_W,m_W\right) &=& \frac{2 \sqrt{4 \lambda-1} \arctan \left(\sqrt{\frac{1}{4 \lambda-1}}\right)-1+\log \left(\frac{m_{W}^{2}}{M_{R}^{2}}\right)}{M_{R}^{2}} \,,  \nonumber\\
C_{0} \left(p_2,p_1,m_W,M_R,M_R\right)&=& -\frac{1}{M_{R}^{2}} \,, \nonumber\\
C_{0} \left(0,p_{lep},M_{R},M_{R},m_W\right) &=& -\frac{1}{M_{R}^{2}}  \,,\nonumber\\
\tilde{C}_{0} \left(p_2,p_1,m_W,M_R,0\right) &=& \tilde{C}_{0} \left(p_2,p_1,m_W,0,M_R\right) \nonumber\\
&=& \Delta 
+1 -\log \Big(\frac{M_{R}^{2}}{\mu^2}\Big)  + \frac{m_{W}^{2} \log \left(\frac{m_{W}^{2}}{M_{R}^{2}}\right)}{M_{R}^{2}}+\frac{m_{H}^{2}}{2 M_{R}^{2}} \,,  \nonumber\\
\tilde{D}_{0} \left(p_2,0,p_1,m_W,0,M_{R},M_{R}\right) &=& \tilde{D}_{0} \left(p_2,p_1,0,m_W,0,M_{R},M_{R}\right) = -\frac{1}{M_{R}^{2}}  \,,\nonumber\\
\tilde{D}_{0} \left(p_2,0,p_1,m_W,M_{R},M_{R},0\right) &=& \tilde{D}_{0} \left(p_2,p_1,0,m_W,M_{R},M_{R},0\right) = -\frac{1}{M_{R}^{2}} \,, \nonumber\\
B_{1} \left(p,M_R,m_W\right) &=& -\frac{\Delta }{2}-\frac{3}{4}+\frac12\log\Big(\frac{M_{R}^{2}}{\mu^2}\Big)-\frac{m_{W}^{2} \left(2 \log \left(\frac{m_{W}^{2}}{M_{R}^{2}}\right)+1\right)}{2 M_{R}^{2}}  -\frac{p^2}{3M_R^2}\,, \nonumber\\
C_{11} \left(p_2,p_1,m_W,0,M_R\right) &=& \frac{1-\log \left(\frac{m_{W}^{2}}{M_{R}^{2}}\right)}{2 M_{R}^{2}}  \,, \nonumber\\
C_{12} \left(p_2,p_1,m_W,0,M_R\right) &=& \frac{1}{2 M_{R}^{2}}  \,, \nonumber\\
C_{11} \left(p_2,p_1,m_W,M_R,0\right) &=& \frac{1-\log \left(\frac{m_{W}^{2}}{M_{R}^{2}}\right)}{2 M_{R}^{2}}   \,,\nonumber\\
C_{12} \left(p_2,p_1,m_W,M_R,0\right) &=& -\frac{\log \left(\frac{m_{W}^{2}}{M_{R}^{2}}\right)}{2 M_{R}^{2}}   \,,\nonumber\\
C_{11} \left(p_2,p_1,M_R,m_W,m_W\right) &=& 2 C_{12} \left(p_2,p_1,M_R,m_W,m_W\right)  \nonumber\\
&=& -\frac{4 \sqrt{4 \lambda-1} \arctan \left(\sqrt{\frac{1}{4 \lambda-1}}\right)+2 \log \left(\frac{m_{W}^{2}}{M_{R}^{2}}\right)-1}{2 M_{R}^{2}} \,,  \nonumber\\
C_{11} \left(p_2,p_1,m_W,M_R,M_R\right) &=& 2 C_{12} \left(p_2,p_1,m_W,M_R,M_R\right) = \frac{1}{2M_{R}^{2}}\,,  \nonumber\\
C_{12} \left(0,p_{lep},0,M_R,m_W\right)&=& \frac{-\log \left(\frac{m_{W}^{2}}{M_{R}^{2}}\right)-1}{2 M_{R}^{2}}  \,,\nonumber\\
C_{12} \left(0,p_{lep},M_R,M_R,m_W\right)&=& \frac{1}{2 M_{R}^{2}}  \,,\nonumber\\
D_{12} \left(0,p_2,p_1,0,M_R,m_W,m_W\right) &=& 2 D_{13} \left(0,p_2,p_1,0,M_R,m_W,m_W\right)   \nonumber\\
&=& \frac{2\left( -4 \lambda \arctan^2 \left(\sqrt{\frac{1}{4 \lambda-1}}\right)+2 \sqrt{4 \lambda-1} \arctan \left(\sqrt{\frac{1}{4 \lambda-1}}\right)-1 \right)}{ M_{R}^{2} m_{H}^{2}} \, . \nonumber\\
\label{allfloops}
\ea
 %%%%%%%%%%%%%
where we have used the usual definitions in dimensional regularization, $\Delta= 2/\epsilon-\gamma_E +{\rm Log}(4\pi)$ with $D=4-\epsilon$ and $\mu$ the usual scale,  and we have denoted the mass ratio $\lambda=\frac{m_{W}^{2}}{m_{H}^{2}}$ to shorten the result.
%%%%%%%%%%%%%%%%%%%%%%%%%%%%%%%%%%%%%%%%%%%%%%%

Taking into account the formulas above in eq.~(\ref{allfloops}),  plugging them into the results of the form factors in the Appendix \ref{app:FormFactors}, neglecting the tiny terms with lepton masses, and pairing diagrams conveniently, we finally get the results for the dominant terms of the various type diagrams (i) of the MIA form factors valid in the large $M_R\gg v$ regime:
\ba
F_{L}^{(1)} &=& \frac{1}{32 \pi^{2}} \frac{m_{k}}{m_{W}} \Bigg[ \left(Y_{\nu} Y_{\nu}^{\dagger}\right)^{km} \left( \Delta +1 -\log (\frac{M_{R}^{2}}{\mu^2}) + \frac{m_{W}^{2} \log \left(\frac{m_{W}^{2}}{M_{R}^{2}}\right)}{M_{R}^{2}}+\frac{m_{H}^{2}}{2 M_{R}^{2}} \right)   \nonumber\\
&& -\frac{5}{2}\frac{v^{2}}{M_{R}^{2}}\left(Y_{\nu} Y_{\nu}^{\dagger} Y_{\nu} Y_{\nu}^{\dagger} \right)^{km}  \Bigg] \,,\nonumber\\
F_{L}^{(2)} &=& -\frac{1}{32 \pi^{2}} \frac{m_{k}}{m_{W}}  \left(Y_{\nu} Y_{\nu}^{\dagger}\right)^{km} \frac{m_{W}^{2}}{M_{R}^{2}} \left( 1+\log \left(\frac{m_{W}^{2}}{M_{R}^{2}}\right) \right)\,,   \nonumber\\
F_{L}^{(3)} &=& \frac{1}{8 \pi^{2}} \frac{m_{k}}{m_{W}} \left(Y_{\nu} Y_{\nu}^{\dagger} \right)^{km} \frac{\lambda m_{W}^{2}}{M_{R}^{2}} \left( -4\lambda \arctan^2 \left(\frac{1}{\sqrt{4\lambda-1}}\right) \right. \nonumber\\
&& \left. +2\sqrt{4\lambda-1}\arctan \left(\frac{1}{\sqrt{4\lambda-1}}\right) -1  \right) \,,\nonumber\\
F_{L}^{(4+5)} &=& \frac{1}{32 \pi^{2}} \frac{m_{k}}{m_{W}} \left(Y_{\nu} Y_{\nu}^{\dagger} \right)^{km} \frac{m_{W}^{2}}{M_{R}^{2}} \left( 8\lambda \arctan^2 \left(\frac{1}{\sqrt{4\lambda-1}}\right) \right. \nonumber\\
&& \left. -2\sqrt{4\lambda-1}\arctan \left(\frac{1}{\sqrt{4\lambda-1}}\right) +\frac{1}{2} +\log \left(\frac{m_{W}^{2}}{M_{R}^{2}}\right)  \right)\,, \nonumber\\
F_{L}^{(6)} &=& \frac{1}{32 \pi^{2}} \frac{m_{k}}{m_{W}} \left(Y_{\nu} Y_{\nu}^{\dagger} \right)^{km} \frac{m_{H}^{2}}{M_{R}^{2}} \left( \sqrt{4\lambda-1}\arctan \left(\frac{1}{\sqrt{4\lambda-1}}\right) -\frac{3}{4} +\frac{\log \left(\frac{m_{W}^{2}}{M_{R}^{2}}\right)}{2}  \right) \,, \nonumber\\
F_{L}^{(7+9)} &=& 0 \,,\nonumber\\
F_{L}^{(8+10)} &=& -\frac{1}{32 \pi^{2}} \frac{m_{k}}{m_{W}} \Bigg[ \left(Y_{\nu} Y_{\nu}^{\dagger}\right)^{km} \left( \Delta +1 -
\log (\frac{M_{R}^{2}}{\mu^2}) + \frac{m_{W}^{2} \log \left(\frac{m_{W}^{2}}{M_{R}^{2}}\right)}{M_{R}^{2}} \right)   \nonumber\\
&& -\frac{v^{2}}{M_{R}^{2}}\left(Y_{\nu} Y_{\nu}^{\dagger} Y_{\nu} Y_{\nu}^{\dagger} \right)^{km}  \Bigg] \, .  
\label{FLsimple_totdom}
\ea
And similar formulas can be obtained for the $F_R$ form factors.
Notice that in the results above we have included all the relevant contributions, i.e. up  to ${\cal O}(Y_\nu^2+Y_\nu^4)$  and it turns out, as announced in the text, that they are just the diagrams (1)+(8)+(10) that provide contributions of  
${\cal O}(Y_\nu^4)$ with a $v^2/M_R^2$ dependence. The other diagrams will also give  ${\cal O}(Y_\nu^4)$ contributions but they will be suppressed since they go with a $v^4/M_R^4$ dependence, and we do not keep  these small contributions in our expansions.
%%%%%%%%%%%%%%%%%%%%%%%%%%%%%%%%
%%%%%%%%%%%%%%%%%%%%%%%%%%%%%%%%%%%%%%
\section{Appendix: MIA Form Factors (unitary gauge)} 
\label{app:othergauge}
In order to check the gauge invariance of our results for the LFVHD form factors (and therefore the partial width) that we have computed in the MIA  by using the Feynman-'t Hooft gauge, we are going to present here the computation of these same form factors but using a different gauge choice, in particular the unitary gauge (UG).   We will demonstrate that when computing the MIA form factor $F_L$ to ${\cal O}(Y_{\nu}^{2}+Y_{\nu}^{4})$ we get the same result as in eq.~(\ref{FLtotdom}). A similar demonstration can be done for $F_R$ but we do not include it here for shortness. For this exercise, we ignore the tiny terms suppressed by factors of the lepton masses as we did in eq.~(\ref{FLtotdom}).

\begin{figure}[t!]
\begin{center}
\includegraphics[scale=0.8]{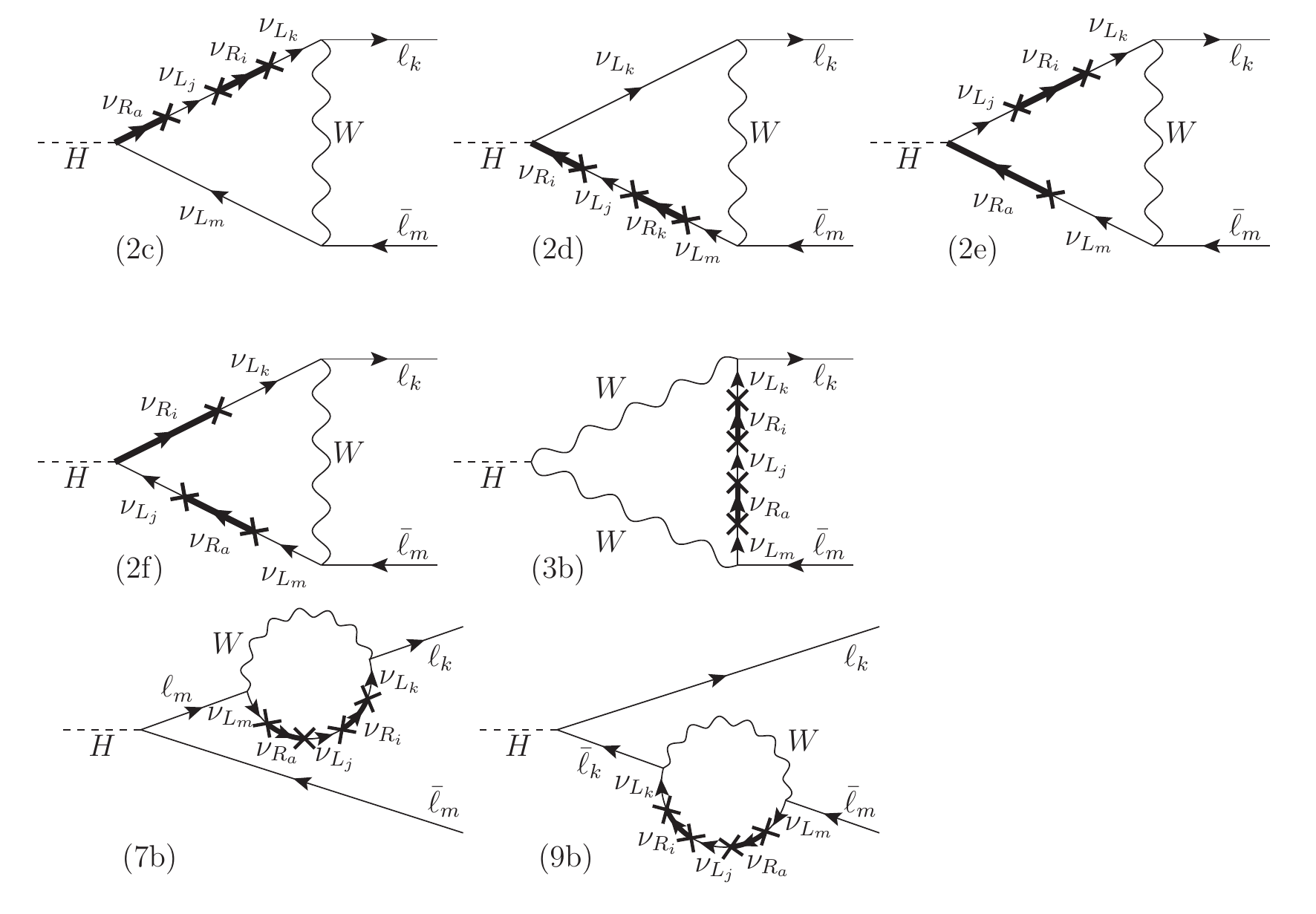}
\caption{Relevant diagrams for the form factors to ${\cal O}(Y_\nu^4)$ in the unitary gauge}
\label{diagsUG}
\end{center}
\end{figure}

First, we list the relevant one-loop diagrams contributing to the form factor $F_L$ in the UG. Since, in this gauge there are not Goldstone bosons, there will be just diagrams of type: (2), (3), (7) and (9).  Generically, each of these diagrams will get contributions of  ${\cal O}(Y_{\nu}^{2})$ and  ${\cal O}(Y_{\nu}^{4})$. Second, we write the propagator of the $W$ gauge boson in the UG, $P_W^{\rm UG}$,  by splitting it into two parts, $P_W^{a}$ and $P_W^{b}$:
\be
P_W^{\rm UG}=P_W^{a}+P_W^{b}=-\frac{i g_{\mu \nu}}{p^{2}-m_{W}^{2}} +\frac{i p_{\mu}p_{\nu}}{m_{W}^{2}(p^{2}-m_{W}^{2})} \, ,
\ee
such that, $P_W^{a}$ coincides with the $W$ propagator in the Feynman-'t Hooft gauge. Then, each diagram of type (i), i=2,3,7,9,  will receive  three kind of contributions: 1) from the part $P_W^{a}$ one gets the same contributions to 
${\cal O}(Y_{\nu}^{2})$ as those we got  in the Feynman-'t Hooft gauge from the five diagrams (2a), (2b), (3a), (7a) and (9a)  in figures~\ref{diagsvertexY2} and \ref{diagslegY2};  2) new contributions to ${\cal O}(Y_{\nu}^{2})$ that come from considering the new propagator term $P_W^b$ in these same diagrams (2a), (2b), (3a), (7a) and (9a) ; 3) contributions to ${\cal O}(Y_{\nu}^{4})$ that come from new diagrams which were not relevant in the Feynman-'t Hooft gauge, but they are relevant in the UG. By relevant we mean leading to dominant ${\cal O}(M_R^{-2})$ contributions in the large $M_R$ expansion. These new diagrams 
contributing to order ${\cal O}(Y_{\nu}^{4})$ in the UG are the seven diagrams shown in figure~\ref{diagsUG}. Thus, we get in total twelve one-loop diagrams contributing in the UG: (2a), (2b), (2c), (2d), (2e), (2f), (3a), (3b), (7a), (7b), (9a) and (9b). 

Next we present the results in the UG for each type of diagram (i), specifying the various contributions explained above, which for clarity we present correspondingly ordered in three lines,  the first line is for kind 1), the second line is for kind 2) and the third line is for kind 3). 
The UG $F_L$ form factors to  ${\cal O}(Y_{\nu}^{2}+Y_{\nu}^{4})$ that we get are, as follows:

 \ba
F_{L}^{\rm UG (2)} &=& -\frac{1}{16 \pi^{2}} m_{k}m_{W} \left(Y_{\nu} Y_{\nu}^{\dagger} \right)^{km} \left( (C_{0}+C_{11}-C_{12})_{(2a)} +(C_{11}-C_{12})_{(2b)} \right) \nonumber \\
&& +\frac{1}{32 \pi^{2}} \frac{m_{k}}{m_{W}} \left[ \left(Y_{\nu} Y_{\nu}^{\dagger} \right)^{km} \left( (\tilde{C}_{0})_{(2a)}-(B_{1})_{(2b)} \right) \right. \nonumber\\
&& \left. +\left(Y_{\nu} Y_{\nu}^{\dagger} Y_{\nu} Y_{\nu}^{\dagger} \right)^{km} v^{2} \left( -(C_{11})_{(2c)} +(\tilde{D}_{0})_{(2d)} +(\tilde{D}_{0}-(C_{11}-C_{12}))_{(2e)} -(C_{11}-C_{12})_{(2f)} \right) \right] \,, \nonumber \\  
F_{L}^{\rm UG(3)} &=& \frac{1}{8 \pi^{2}} m_{k}m_{W}^{3} \left(Y_{\nu} Y_{\nu}^{\dagger} \right)^{km} ( D_{12}-D_{13} )_{(3a)} \nonumber \\
  & & -\frac{1}{32 \pi^{2}} \left(Y_{\nu} Y_{\nu}^{\dagger} \right)^{km} \frac{m_{k}}{m_{W}} \left[ 2B_{0}+B_{1} -(2m_{W}^{2}+m_{H}^{2})(C_{0}+C_{11}-C_{12}) +2m_{W}^{2}m_{H}^{2}D_{13} \right]_{(3a)} \nonumber\\
&& -\frac{1}{32 \pi^{2}} \left(Y_{\nu} Y_{\nu}^{\dagger} Y_{\nu} Y_{\nu}^{\dagger} \right)^{km} \frac{m_{k}}{m_{W}} v^{2} \left[ 2C_{0}+C_{12} \right]_{(3b)} \,, \nonumber \\
F_{L}^{\rm UG(7)} &=& \frac{1}{16 \pi^{2}} m_{k}m_{W} \frac{m_{m}^{2}}{m_{k}^{2}-m_{m}^{2}}  \left(Y_{\nu} Y_{\nu}^{\dagger} \right)^{km} ( C_{12} )_{(7a)}  \,,\nonumber \\
  && \frac{1}{32 \pi^{2}} \frac{m_{k}m_{m}^{2}}{m_{W}(m_{k}^{2}-m_{m}^{2})} \left[ \left(Y_{\nu} Y_{\nu}^{\dagger} \right)^{km}( 2B_{0}+B_{1} )_{(7a)} \right. \nonumber\\
&& \left. +\left(Y_{\nu} Y_{\nu}^{\dagger} Y_{\nu} Y_{\nu}^{\dagger} \right)^{km} v^{2}(2C_{0}+C_{12})_{(7b)} \right]  \nonumber \\
F_{L}^{\rm UG(9)} &=& -\frac{1}{16 \pi^{2}} m_{k}m_{W} \frac{m_{m}^{2}}{m_{k}^{2}-m_{m}^{2}}  \left(Y_{\nu} Y_{\nu}^{\dagger} \right)^{km} ( C_{12} )_{(9a)} \,,\nonumber \\ 
  && -\frac{1}{32 \pi^{2}} \frac{m_{k}m_{m}^{2}}{m_{W}(m_{k}^{2}-m_{m}^{2})} \left[ \left(Y_{\nu} Y_{\nu}^{\dagger} \right)^{km}( 2B_{0}+B_{1} )_{(9a)} \right. \nonumber\\
&& \left. +\left(Y_{\nu} Y_{\nu}^{\dagger} Y_{\nu} Y_{\nu}^{\dagger} \right)^{km} v^{2}(2C_{0}+C_{12})_{(9b)} \right] \, ,
\label{FL_Uxi1}
\ea
where the arguments of the one-loop functions are
\be\begin{array}{rll}
 \tilde{C}_{0}, C_{i} =& \tilde{C}_{0}, C_{i}(p_{2},p_{1},m_{W},0,M_{R})  & \text{in } (2a) \nonumber \\
 B_{i} = & B_{i}(p_{lep},m_{W},M_{R})  & \text{in } (2b)  \nonumber \\
 C_{i} = & C_{i}(p_{2},p_{1},m_{W},M_{R},0)  & \text{in } (2b) \nonumber \\
 C_{i} =  & C_{i}(p_{lep},0,m_{W},M_{R},M_{R})  & \text{in } (2c)  \nonumber \\
 \tilde{D}_{0} = & \tilde{D}_{0}(p_{2},p_{1},0,m_{W},0,M_{R},M_{R})  & \text{in } (2d)  \nonumber \\
 C_{i} = &  C_{i}(p_{2},p_{1},m_{W},M_R,M_{R}) & \text{in } (2e),(2f)  \nonumber \\
 \tilde{D}_{0} =&  \tilde{D}_{0}(p_{2},0,p_{1},m_{W},0,M_{R},M_{R})  & \text{in } (2e)  \nonumber \\
 B_{i} = & B_{i}(p_{lep},M_{R},m_{W})  & \text{in } (3a), (7a), (9a)  \nonumber \\
 C_{i} = & C_{i}(p_{2},p_{1},M_{R},m_{W},m_{W})  & \text{in } (3a) \nonumber \\
 D_{i} = & D_{i}(0,p_{2},p_{1},0,M_{R},m_{W},m_{W})  & \text{in } (3a)  \nonumber \\
 C_{i} = & C_{i}(0,p_{lep},M_{R},M_R,m_{W})  & \text{in } (3b), (7b), (9b) \nonumber \\
 C_{i} = & C_{i}(0,p_{lep},0,M_R,m_{W})  & \text{in } (7a), (9a) \, . \nonumber \\
\label{argfloops_U}
\end{array}\ee
The comparison of the previous results  with that in  eq.~(\ref{FLtotdom}) then goes as follows.  First, it is clear from the above results, that once again the contributions from diagrams (7) and (9) cancel out fully, as it happened in the Feynman-'t Hooft gauge.  Therefore, $F_{L}^{\rm UG} =F_{L}^{\rm UG(2)}+ F_{L}^{\rm UG(3)}$. Then, the first line in $F_{L}^{\rm UG(2)}$ and the first line in $F_{L}^{\rm UG(3)}$ match correspondingly with the contributions from (2) and (3) in the Feynman-'t Hooft gauge.  Next, by using the relation,  
\be
 B_{0}(p_{lep},M_{R},m_{W})+B_{1}(p_{lep},M_{R},m_{W}) +B_1(p_{lep},m_{W},M_{R})= 0  \,,
\ee
we get that the sum of the second line in $F_{L}^{\rm UG(2)}$ and the second line in $F_{L}^{\rm UG(3)}$ gives exactly the contributions to 
${\cal O}(Y_\nu^2)$ from (1)+(8)+(10)+(4)+(5)+(6) in the Feynman-'t Hooft gauge. Finally, by using the relation 
\be
 C_{11}(p_{lep},0,m_{W},M_{R},M_{R})+(C_{0}+C_{12})(0,p_{lep},M_{R},M_{R},m_{W}) = 0  \, ,
\label{iden_compU}
\ee
we get that the sum of the third line in $F_{L}^{\rm UG(2)}$ and the third line in $F_{L}^{\rm UG(3)}$ gives exactly the contributions to  ${\cal O}(Y_\nu^4)$ from (1)+(8)+(10). Therefore, in summary, we get the identity of the total result for $F_L$ computed in both gauges, leading to the gauge invariant result of eq.~(\ref{FLtotdom}).

%%%%%%%%%%%%%%%%%%%%%%%%%%%%%%%%%%%%%%

\end{appendix}

%%%%%%%%%%%%%%%%%%%%%%%%%%%%%%%%

\bibliography{bibliography}

\end{document}